\documentclass[11pt]{article}
\pdftrailerid{}
\input{glyphtounicode}
\usepackage[letterpaper,margin=1in]{geometry}
\usepackage[T1]{fontenc}
\usepackage[utf8]{inputenc}
\usepackage{mathptmx}
\usepackage[scaled=0.92]{helvet}
\usepackage{courier}
\usepackage{textcomp}
\usepackage{newunicodechar}
\newunicodechar{→}{\ensuremath{\rightarrow}}
\usepackage{microtype}
\usepackage{graphicx}
\usepackage{booktabs}
\usepackage{tabularx}
\usepackage{enumitem}
\usepackage{fvextra}
\usepackage{upquote}
\usepackage{caption}
\DeclareCaptionLabelSeparator{dash}{ - }
\usepackage[hidelinks,pdfusetitle=false]{hyperref}
\hypersetup{pdftitle={},pdfauthor={},pdfsubject={},pdfkeywords={},pdfcreator={},pdfproducer={}}
\expandafter\def\expandafter\UrlBreaks\expandafter{\UrlBreaks\do\-}
\usepackage[parfill]{parskip}
\usepackage{needspace}
\usepackage{xcolor}
\usepackage{placeins}
\makeatletter
\newcommand\nohy[1]{{\language\l@nohyphenation #1}}
\makeatother

\makeatletter
\renewcommand\@seccntformat[1]{\@ifundefined{seccnt@#1}{\csname the#1\endcsname\quad}{\csname seccnt@#1\endcsname}}
\def\seccnt@section{\thesection.\ }
\def\seccnt@subsection{\thesubsection\ }
\makeatother
\setlist{itemsep=0.25em,topsep=0.4em}
\definecolor{codecolA}{rgb}{0.000,0.439,0.125}
\definecolor{codecolB}{rgb}{0.098,0.090,0.486}
\definecolor{codecolC}{rgb}{0.376,0.627,0.690}
\definecolor{codecolD}{rgb}{1.000,0.000,0.000}
\definecolor{codecolE}{rgb}{0.024,0.157,0.494}
\definecolor{codecolF}{rgb}{0.400,0.400,0.400}
\definecolor{codecolG}{rgb}{0.251,0.439,0.627}
\definecolor{codecolH}{rgb}{0.000,0.502,0.000}

\title{Where AI Vulnerability Discovery Harnesses Go Quiet: Structural False Negatives and Hidden Vulnerabilities}
\author{Gadi Evron\\ \nohy{gadi@knostic.ai}}
\makeatletter
\def\@maketitle{\newpage\null\vskip 2em\begin{center}\let\footnote\thanks{\LARGE \@title \par}\vskip 1.5em{\large\lineskip .5em\begin{tabular}[t]{c}\@author\end{tabular}\par}\end{center}\par\vskip 1.5em}
\makeatother
\begin{document}
\maketitle
\begin{abstract}
LLM-based vulnerability discovery harnesses can miss real, reachable vulnerabilities because of how the harness around the model is built. Steps that select code, filter it, and store and merge findings can drop a vulnerability before or after the model runs, independently of the model’s capability, and the report usually gives no sign that this happened. These drops come from engineering choices, and errors. They leave vulnerabilities undetected, and an attacker who knows the harness’s rules can also use them to hide a vulnerability on purpose. Understanding these drops shows how to build harnesses that fix or disclose them and are harder to evade.

Starting from the two harnesses we maintain, Knostic’s \nohy{OpenAnt} and raptor, and then examining 10 additional open-source harnesses, we mapped 13 points where such drops occur and organized them into a taxonomy. Using each harness’s own code, we show a single file that six harnesses dropped before their model saw its code, five of them without notice. We propose engineering failure patterns mapped to these drops, and relate them to Ptacek and Newsham’s work on intrusion detection evasion. The paper also covers content-steered verdicts and prompt injection, and gives design principles and tests for harnesses that report accurately what they analyzed. The reliability findings likely apply to other LLM harnesses, beyond vulnerability research and application security, and the evasion findings apply wherever an attacker can write or influence the code being checked.

We also argue that harnesses built to be descriptive can, in effect, be prescriptive. Rules written to describe typical code and typical model replies end up deciding what is analyzed, and anything that does not conform is treated as absent or as “no vulnerability.”
\end{abstract}

\section{Introduction}

A harness report with no findings can mean three things: the code is clean, the model analyzed the vulnerable code and missed the vulnerability, or the vulnerability never reached the model or was removed after it. The report usually looks the same in all three cases.

This paper is about the third case. We map the places where a harness structurally misses a vulnerability because a deterministic stage in the pipeline drops it. Sometimes the code is dropped before anything analyzes it (a file not selected, a macro the parser will not read). Sometimes a real finding is formed, and a later stage silently rejects it, suppresses it, overwrites it, or serves a stale “clean” from a cache.

We first found these drops while hardening the two harnesses we maintain, Knostic’s \nohy{OpenAnt} and the open-source raptor. We then made the problem a research project: we examined how deep it goes in the harnesses we maintain and how widely it appears in other open-source harnesses. The two harnesses we maintain, \nohy{OpenAnt} and raptor, were examined most closely, and are intentionally used for many of the examples in this text.

Two lines of earlier work frame what we found. Ptacek and Newsham (1998) showed that a network intrusion detection system can be evaded when it processes traffic differently from the host it protects. Many of the drops in this paper come from the same gap, with the harness as the monitor and the runtime as the host. Ptacek and Newsham studied the gap as a means of attack, and here it also causes misses with no attacker present. Other drops come from a harness losing results it already had, and there the relevant rule is the fail-safe default (Saltzer \& Schroeder, 1975), applied to results: a failure should not be reported as a pass.

These drops matter in two ways. In ordinary use, they leave real vulnerabilities undetected while the report suggests the code was checked. When the code’s author is hostile, they offer a way to hide a vulnerability on purpose. Our aims follow from this. The first is to show where and how harnesses drop vulnerabilities, including how an attacker can cause it. The second is to derive principles for building resilient harnesses, meaning harnesses that keep uncertain code in scope, report what they skipped, and are harder to evade.

Most of the drops happen in deterministic steps added to control cost and false positives. We argue that harnesses built to be descriptive can, in effect, be prescriptive.\footnote{We use descriptive and prescriptive in their general sense: a description states how things are, and a prescription states how they must be. In software architecture, Taylor, Medvidovic and Dashofy (2010) use the terms for as-implemented and as-intended architecture, which is a different distinction.} A rule written to describe typical code, such as “Python files end in .py,” or a typical model reply, such as “the answer begins with TRUE,” comes to define what is analyzed and how replies are read. Code or replies that do not conform are treated as absent or as “no vulnerability,” and these limits are often not considered when the rules are written.

Published evaluations of these harnesses measure precision, exploitability and cost, but not recall lost in the harness’s own steps (Korda \& Evron, 2026; Guo et al., 2025; Li et al., 2025).

Our contributions are:

\begin{itemize}

\item A catalogue of 13 drop mechanisms observed in 12 harnesses, with the harnesses that show each one, and a definition of when a drop counts as silent (Sections 2 and 3).

\item Examples in 12 harnesses, each traced to source code or shipped instructions at a specific commit (Section 4).

\item Demonstrations of hiding a vulnerability from one harness, from one harness with three stacked placements, and from several harnesses at once (Sections 5 to 7).

\item An initial analysis of why harnesses miss vulnerabilities, organized by the design principle each drop violates (Section 8).

\item Recommendations for building, testing and using harnesses, each tied to the findings (Section 9).

\end{itemize}

Section 10 lists limitations, Section 11 describes our disclosure, and Section 12 concludes.

\subsection{The harnesses we studied}

\textbf{They are all open-source tools marketed as AI- or LLM-powered code security scanners:} most wrap a language model around some static-analysis machinery. They do \textbf{not} all work alike, and fall into three shapes: harnesses that run their own analysis (build their own call graph), harnesses that drive an engine like Semgrep or \nohy{CodeQL,} and harnesses that are mostly models with only thin plumbing around them.

Each harness is built from some of the pipeline stages shown in Section 3, and a real bug can fall out at any stage a harness runs, before or after the model.

As mentioned, most of the examples in this text are about Knostic’s \nohy{OpenAnt,} and raptor, as the authors are maintainers of the projects, and the work started as a study of these issues for these harnesses specifically.

\begin{center}\small
\begin{tabularx}{0.846\textwidth}{>{\raggedright\arraybackslash\hsize=1.000\hsize}X>{\raggedright\arraybackslash\hsize=1.000\hsize}X>{\raggedright\arraybackslash\hsize=1.000\hsize}X>{\raggedright\arraybackslash\hsize=1.000\hsize}X>{\raggedright\arraybackslash\hsize=1.000\hsize}X}
\toprule
\textbf{harness} & \textbf{vendor} & \textbf{what it is} & \textbf{built on} & \textbf{drop classes} \\
\midrule
\nohy{OpenAnt} & Knostic & own-analysis harness, six languages & own parsers and call graph (tree-sitter for Rust) & 1, 3, 4, 6, 11 (d), 12, 13 (d) \\
raptor & open-source project & harness that drives Semgrep and \nohy{CodeQL} & Semgrep, \nohy{CodeQL} & 7, 8 (c), 9, 10 \\
ai-deep-sast & Cisco & Semgrep wrapper & Semgrep & 1, 8 \\
Codex Security & \nohy{OpenAI} & pure-LLM agentic reviewer with file tools & LLM, no wrapped engine & 1 \\
ccsr (Claude Code Security Review) & Anthropic & \nohy{GitHub} Action, model reviews each PR & model with file tools & 6 (c), 11 \\
\nohy{RepoAudit} & open-source project & LLM-guided dataflow harness & tree-sitter call graph, LLM & 1, 2, 7 \\
iris & open-source project & LLM-labeled taint harness (Java) & \nohy{CodeQL} & 4, 7, 12 \\
\nohy{ToB} (skills) & Trail of Bits & \nohy{CodeQL} and Semgrep wrapper skills & \nohy{CodeQL,} Semgrep & 1, 11, 12 \\
deepsec & Vercel & harness with diff mode and finding merge & not stated & 1, 5, 13 \\
VVAH (Visa Vulnerability Agentic Harness) & Visa & harness reading the whole tree & specialist analyzers & 9, 11 (d) \\
\nohy{VulnHunter} & Capital One & batch harness with -{}-resume & not stated & 10 \\
ai-sast & Rivian & pure-LLM, \nohy{per-file/diff} harness & LLM, no wrapped engine & 7, 11 \\
\bottomrule
\end{tabularx}
\end{center}

Plain numbers are drops we found silent under Rules 1 and 2 (Section 2). A (d) marks a drop the harness discloses, and a (c) marks a drop that fires only in a fallback or opt-in mode.

(Two harnesses share the “ai-” naming, Cisco’s ai-deep-sast and Rivian’s ai-sast. They are different projects.)

\section{What a false negative is, and the kind we care about}

A \textbf{false negative} is a real vulnerability a harness misses, or, in our case, one it did not report.

As in Section 1, a harness that reports nothing has told you one of three things, and the report looks identical in all three:

\begin{enumerate}[label=\arabic*.,start=1]

\item The code is clean,

\item The model looked at the code and judged it safe (a model miss), or

\item The vulnerability never reached the model or the report. For example, its file was never selected, the call graph refused to connect it, a cache served a stale “nothing here,” or a filter removed it.

\end{enumerate}

Here we make the third case precise.

This example traces the real path the vulnerability takes, from the harness’s entry point, through the stage that drops it, to the sink it would reach, so we can point at exactly where it is lost. \nohy{OpenAnt’s} extension drop sits here on that path (Figure 1):

\begin{quote}\raggedright
\texttt{openant parse} (CLI) → \texttt{scan\_\allowbreak{}repository} → \texttt{parse\_\allowbreak{}repository} → \texttt{detect\_\allowbreak{}languages} (routes by file suffix) → the \nohy{PHP/Python} parser subprocess → \texttt{RepositoryScanner.\allowbreak{}scan\_\allowbreak{}directory} → \texttt{if not is\_\allowbreak{}source\_\allowbreak{}file(entry.\allowbreak{}name): return} - \textbf{the no-extension file exits here, before any parser or sink analysis.}
\end{quote}

\begin{figure}[htbp]
\centering
\includegraphics[width=0.897\textwidth]{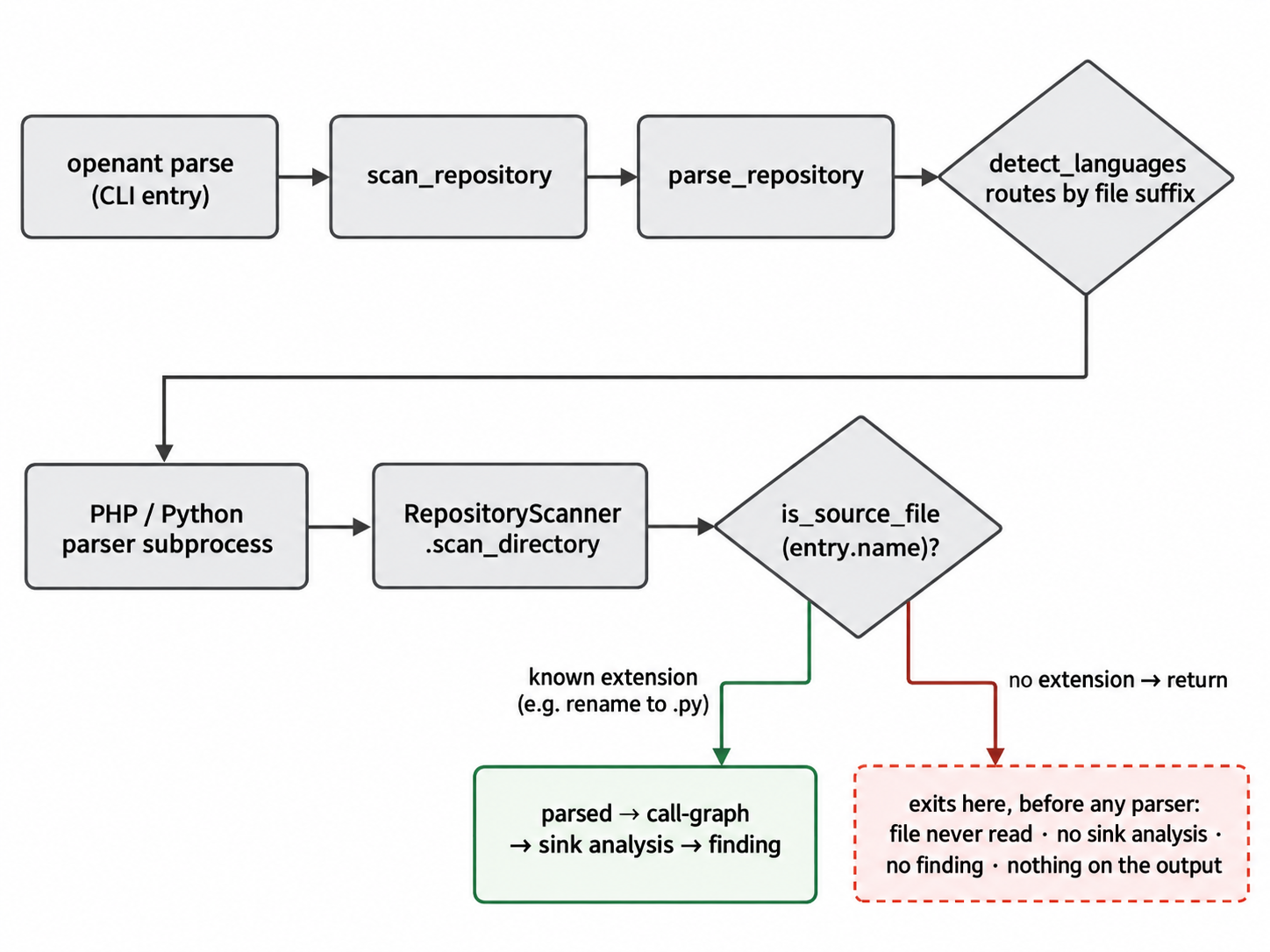}
\caption{\textit{The same trace as a path: \nohy{OpenAnt’s} entry point flows left-to-right until }\textit{\texttt{is\_\allowbreak{}source\_\allowbreak{}file}}\textit{ tests the suffix. A known extension continues to sink analysis (the recover control - rename the file }\textit{\texttt{.py}}\textit{). A no-extension file hits }\textit{\texttt{return}}\textit{ and exits before any parser runs, so no finding is ever formed.}}
\label{fig:1}
\end{figure}

Sometimes no finding was ever formed, because a deterministic stage lost the path before the model saw it. Other times a finding formed and a later deterministic stage dropped it.

We separate a harness that cannot prove a path is reachable (a disclosed limit) from one that silently drops a path that is reachable (the failure we map). Reachable here means the bug can be hit from a live entry point, whatever the harness’s own call graph believes. We also separate a true drop from a disclosed drop: if the harness’s report, or a file it names as output, records an excluded \nohy{tests/} directory with a count, or lists the skipped unit, the coverage loss was announced, and we do not define that as silent.

\subsection{What counts as silent}

Every “silent” claim here rests on one judgment: did the harness tell the user coverage was lost? We answer it the same way for every harness (Figure 2):

\needspace{5\baselineskip}\textbf{Rule 1, where we look (the deliverable)}\par\nopagebreak

We count only what reaches a consumer: the harness’s report, its SARIF, its pull-request comment, its exit status, and any file the harness’s own docs name as a scan output. We do \textbf{not} count console text, stderr, logs, internal state, or \nohy{scratch/intermediate} files - a user reading a report never sees those. (A sidecar file counts only if the report points you to it, one hop away. A file you would find only by knowing the harness’s internals does not count.)

\needspace{5\baselineskip}\textbf{Rule 2, what counts as telling you}\par\nopagebreak

A drop is \textbf{disclosed} (so, \textit{not} silent) if a deliverable either \textbf{(a) names the dropped item} under any label (“rejected,” “pruned,” “skipped,” or “excluded”), or \textbf{(b) carries a \nohy{skip/prune} count that moved because of the drop} \textit{and that count can be tied to the dropped item} (for example VVAH’s \nohy{excluded.dirs:} \{tests: 1\}, when our file is the one thing under \nohy{tests/).} If neither holds, the drop is \textbf{silent}.

Two things deliberately do \textit{not} count: a plain total that merely got smaller (a scan reporting “7 files” instead of “8” has not identified \textit{which} file, or that one was dropped), and a notice that lives only on a non-deliverable surface like stderr. (Throughout, when we say a harness announces, records, or names a drop, we mean it meets Rule 2, so the drop is disclosed.)

\begin{figure}[htbp]
\centering
\includegraphics[width=0.646\textwidth]{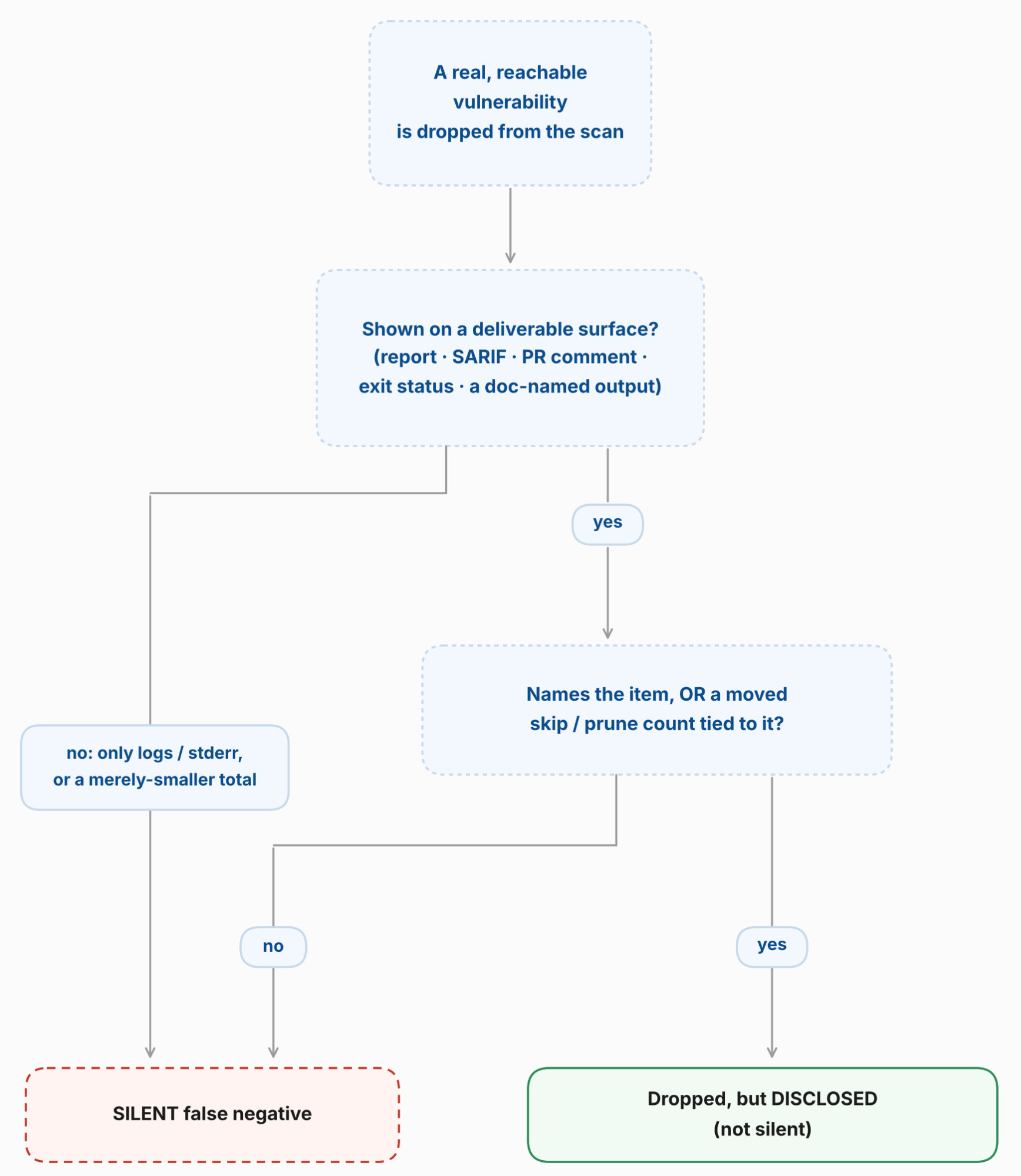}
\caption{\textit{The two rules as one test: Rule 1 (is the drop on a deliverable surface?) then Rule 2 (does it name the item, or carry a moved skip or prune count tied to it?). The four worked examples below are its leaves. ccsr’s .md finding written to \nohy{claudecode-results.\allowbreak{}json,} VVAH’s excluded-directory count and \nohy{OpenAnt’s} \nohy{test\_files\_skipped} count land on the right (disclosed). \nohy{OpenAnt’s} extension drop (only a smaller total, plus a stderr line) and Cisco’s ai-deep-sast \nohy{/} \nohy{RepoAudit} (nowhere - on no deliverable) land on the left (silent).}}
\label{fig:2}
\end{figure}
\FloatBarrier

Four cases from this study show the test in use:

\begin{itemize}

\item \textbf{In the findings file but not the pull-request comment → disclosed, not silent.} Anthropic’s ccsr drops any finding attributed to a \texttt{.md} file. It never appears in the pull-request comment a reviewer reads, but the full finding, labeled “Finding in Markdown documentation file,” is written into the run’s \texttt{claudecode-results.\allowbreak{}json}, which the action’s docs name as its results file. A deliverable names it, so we call this dropped but disclosed. The concern there is a wrong exclusion rather than silence.

\item \textbf{A moved count on a named output → disclosed.} VVAH, scanning a file under \texttt{tests/\allowbreak{}}, writes \nohy{excluded.dirs:} \{tests: 1\} into its report. \nohy{OpenAnt’s} test-path skip writes a \nohy{test\_files\_skipped} count into \texttt{scan\_\allowbreak{}results.\allowbreak{}json}, a documented output. The item is not named, but a count that moved on a deliverable is disclosed.

\item \textbf{Only a smaller total, or only stderr → silent.} \nohy{OpenAnt’s} extension drop leaves no skip counter at all - \texttt{total\_\allowbreak{}files} is simply one lower (a plain total, which Rule 2 does not count), and the one “N skipped” line it prints goes to stderr (which Rule 1 excludes). Nothing on a deliverable names the file or counts its loss.

\item \textbf{Nowhere → silent. }Cisco’s ai-deep-sast and \nohy{RepoAudit} name the dropped file on no deliverable and carry no moved skip count anywhere.

\end{itemize}

So “silent” is a specific claim: \textit{no consumer-facing deliverable told you}, as opposed to “we could not find it mentioned somewhere.”

Where a harness discloses the drop we count it disclosed, even when the disclosure is in a results file and not in the headline report.

\subsection{How we examined the harnesses}

We studied each harness at one pinned commit, by reading its source for every stage of the Section 3 pipeline that it implements. Where a drop depended on runtime behavior, we ran the stage in isolation, without the model, on a planted input and inspected every deliverable the harness produces (Rule 1).

Where an entry in Section 4 says we ran, reproduced or confirmed something, we did so at the cited commit and inspected every deliverable. Every other entry rests on reading the cited source.

Two results depended on a model, both with gpt-6-astra: raptor’s corroboration gate (class 7) was confirmed end to end in six runs, and the planted-comment case in Section 3.2 used three trials at temperature 0.7. The runs date from August to October 2026.

The \nohy{OpenAnt} runs of the entry-point detector (class 4), the Section 6 gate checks and the Section 7.2 selection panel were made at an earlier \nohy{OpenAnt} commit, c81c0e1 (9 July 2026), on a clone pinned there. The \nohy{OpenAnt} citations in Section 4 are at the study pin, b9d5d2f (11 August 2026), except class 4, which is cited at c81c0e1 where it was run, with the pin’s line numbers noted in the entry.

The selection code that the Section 6 checks and the Section 7.2 runs exercised (the extension and test-file checks of the Python and PHP scanners, and the diff filter) is identical at both commits. The entry-point detector changed between them, as the class 4 entry notes.

The Section 5 name-veto run was made at the pin and repeated at c81c0e1 and at the repository head, and the drop reproduces at all three. raptor’s compiler-scan module (class 8) exists only at a later commit, 1b84ec5, and is cited there.

\subsection{Threat model}

A harness can miss a vulnerability in two settings, and this paper covers both. In the first there is no attacker: a drop happens on ordinary code, the report comes back clean, and the vulnerability stays in the product until someone else finds it, either a researcher who discloses it or an attacker who exploits it directly.

In the second the attacker is the code’s author, who can contribute code that the harness will scan, by committing to a repository, opening a pull request, or choosing the names, paths, comments and contents of files in it. The attacker cannot change the harness, its configuration, its cache or its prompts, and cannot see its logs. The attacker can potentially also place a file outside a pull request’s diff, and so outside a diff-mode scan.

Success for the attacker is a report that names no finding for a vulnerability that is live and reachable. The drops are the same in both settings.

\section{The drop points we mapped}

While we do not claim this is an exhaustive list, below are thirteen such drop points - stages where a real, reachable vulnerability can fail to enter the pipeline, or can leave it, often with no notice.

The thirteen are deterministic stages, and many of the examples come from Knostic’s \nohy{OpenAnt} and raptor, the harnesses we maintain.

We present them as mechanisms observed, with the harnesses that show each one named in Section 4, and make no claim about how common they are. After the thirteen we describe two cases in which the attacker works on the harness’s own model, and we close with the drop points we saw but did not map.

\subsection{The thirteen}

\begin{enumerate}[label=\arabic*.,start=1]

\item \textbf{Unrecognized File Types} - The harness ignores files that have no extension, or an extension it does not recognize.

The harness, and not the operating system, decides which files to read by checking each file’s extension against a fixed list, often to pick a language. We saw two forms: a full-repo scan that never reads a file with no extension or an unusual one, and a pull-request scan that drops any changed file whose extension is not on its list.

\item \textbf{Argument Count Mismatch} - The harness only links a call to a function when the number of arguments matches exactly, so valid calls with optional or extra arguments are not followed.

To link a call to the function it calls, the harness uses a shortcut: the number of arguments must equal the number of parameters. Calls to functions with optional or variable arguments can fail this test even though they are valid. Those calls are left out, so the harness never follows data from the caller into the function.

\item \textbf{Built-in Name Conflict} - If a custom function has the same name as a built-in one, calls to it are mistaken for the built-in, so the custom function looks unused and is never checked.

The harness checks each call’s name against a list of built-in functions before working out which function the call actually reaches. If the project defines its own function with a built-in’s name, calls to it are treated as calls to the built-in. The project’s function then has no callers on record, looks unreachable, and is dropped from analysis.

\item \textbf{Unrecognized Entry Points} - Code started by a framework (like a background task or cloud function handler) is treated as unused, because the harness does not recognize it as a starting point.

The harness finds reachable code by starting from known entry points, such as web routes, and following calls. Code that a framework calls directly, such as a background task, a cloud function handler, or an event listener, is not recognized as an entry point, so it is treated as dead code and dropped, even though it was read.

We saw two causes: a fixed list of entry-point patterns that leaves these out, and a query that shows the model only functions the project’s own tests call, which these handlers are not.

\item \textbf{Overridden Findings} - A correct new finding is thrown away because an older result takes priority. For example, a bug once marked as fixed comes back and stays hidden.

A correct result can be lost when results are stored and merged. This happens in one of two ways: one step saves a result in one field and the next step reads a different field, or an older stored verdict takes priority over a new one.

We have seen only the second in a real harness: when a bug that was marked as fixed or as a false positive comes back, the new finding is merged into the old record as a duplicate, and the old status wins. The finding is left out of the count, the report, and the export, with no warning.

\item \textbf{Scan Limit Cutoff} - The harness stops at a file, size, or time limit and quietly drops whatever is left.

A fixed limit caps how much gets scanned, and whatever falls past it is dropped without notice.

We saw two forms: a cap that sorts code by priority and analyzes only the top N items, and a harness that fetches only the first 100 changed files of a pull request and never asks for the rest (this one matters only in a fallback mode). The same applies to per-item limits, such as a timeout or memory limit that stops analysis partway through.

\item \textbf{Verification Filtering} - Real issues are thrown out at a double-check step, because no other engine confirms them, the model wrongly rejects them, or the model’s answer is badly formatted and counted as a no.

Even after a finding is formed, a later review step can throw it out.

We saw three forms. A rule downgrades any finding not backed by an engine other than the model, even when the model is right. A second model check wrongly decides the code cannot be reached. In three harnesses, code expects the model’s reply in an exact format (such as starting with “TRUE”, or valid JSON) and treats any other or cut-off reply as “no vulnerability”.

\item \textbf{Suppression Comments} - A comment or marker in the code (like \#nosemgrep) hides the issue on that line, and the harness never says anything was hidden.

Many analysis engines let developers silence a warning with a special comment or compiler directive on the same line. When a harness runs such an engine without the option that disables this, anyone who can edit the code, including an attacker, can hide a real vulnerability by adding that marker.

We saw this with a Semgrep-based harness (\# nosemgrep) and with an optional compiler-based scan (compiler suppression markers). The harness reports neither the suppression nor a count of hidden findings.

\item \textbf{Inactive Rules} - Security rules or checks that exist in the code or documentation are never actually run during scans.

A rule or analyzer ships with the harness, and may even be documented, but the scan never runs it, so vulnerabilities only it would catch are never looked for. We saw two forms: custom queries documented for users but never added to the set the scan runs, and a complete analyzer for one vulnerability type that no shipped configuration ever turns on.

\item \textbf{Stale Caching} - The harness reuses old results that said the code was clean, even after the code or the harness itself has changed.

The harness saves results and reuses them when it thinks nothing has changed, but it checks the wrong things.

We saw two forms. First, a cache that never expires, keyed on the exact request sent to the model (which includes the code) but not on the harness’s version: after the harness’s logic changes, unchanged code keeps its old “clean” verdict. Second, a resume option that skips any repository that already has a results file, so a repository that became vulnerable since then keeps its old result.

\item \textbf{Keyword-Based Skipping} - Files are skipped if their path, name, or contents include certain keywords (like log or @generated). This can skip real code by mistake, such as \nohy{login.py.}

Before deep analysis, the harness decides what to skip by simple text matching, without reading the code.

We saw two forms. One matches on the path, so any path containing a word like “log” or “script” is skipped and \nohy{auth/login.py} and \nohy{transcript.go} are never scanned. The other matches on content, so a file’s whole change is dropped if any line contains “@generated”, even when the file is not generated.

\item \textbf{Unparsed Code Blocks} - The harness cannot see inside some code, like macros or code the build never compiles, so any vulnerabilities there are missed.

Before analysis, the harness breaks code into pieces (functions and the calls between them). Code it cannot see inside produces no pieces, and code reached only through it then looks unreachable and is dropped too.

We saw two forms: a parser that treats a macro’s body as an unreadable block, and a harness that captures code only while the project is compiled, so code the build skips is never captured.

\item \textbf{Diff Blindspots} - Scans that only look at added or edited files miss two cases: a deleted file that held a safety check, and an unchanged file made unsafe by a change elsewhere.

In pull-request mode, the harness asks git only for files that were added, modified, renamed, or copied, so deleted files are left out. If a change deletes a file that held a security check, or makes an unchanged file unsafe, the affected code is never on the list and never scanned.

\end{enumerate}

Three of these are a harness’s reachability analysis going wrong after the file was read: Argument Count Mismatch (2, seen in \nohy{RepoAudit),} Built-in Name Conflict (3, \nohy{OpenAnt)} and Unrecognized Entry Points (4, \nohy{OpenAnt} and iris).

In each, the vulnerable code is reachable, but a dropped call edge or an unrecognized entry point makes the harness conclude it is dead and discard it.

That differs from the code never reaching the analyzer at all (never selected, or never parsed into a unit when it hides inside a macro the parser cannot read), and again from a formed finding being filtered out later.

\begin{figure}[htbp]
\centering
\includegraphics[width=0.897\textwidth]{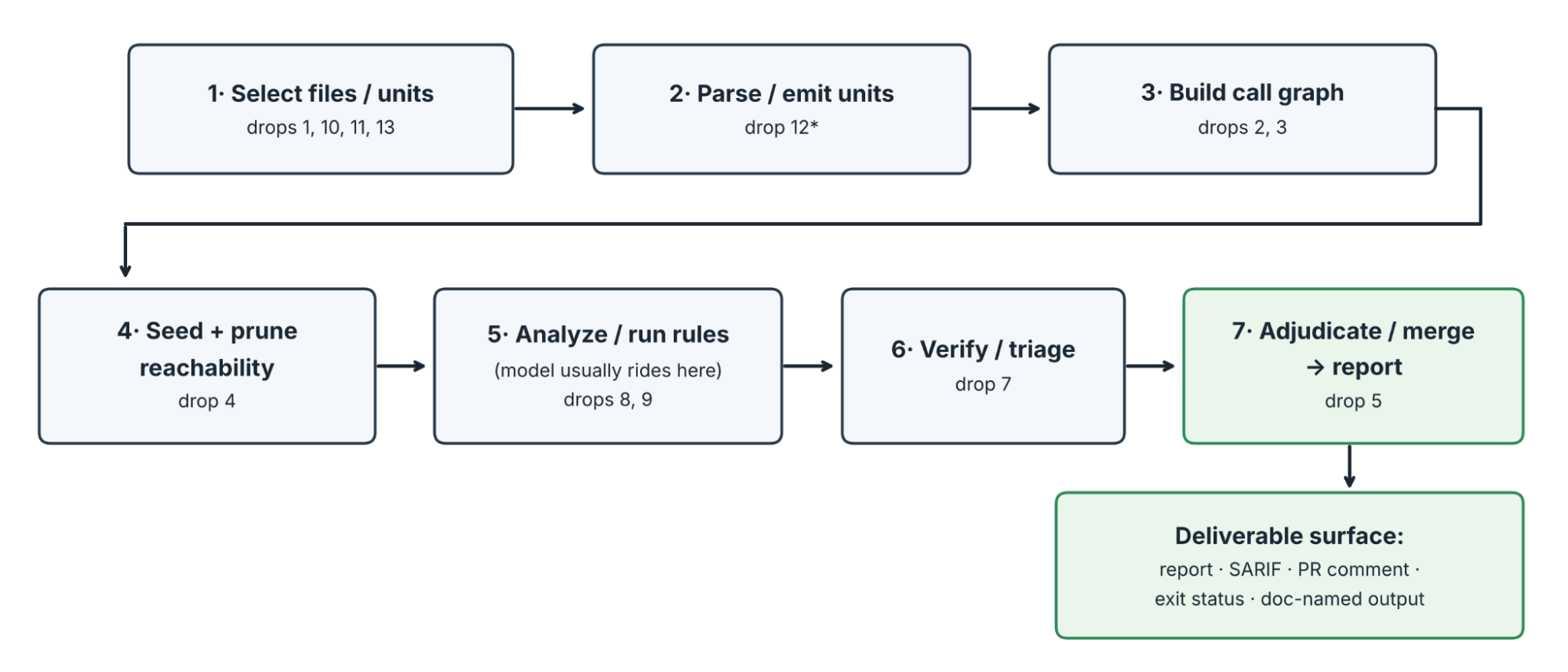}
\caption{\textit{The harness pipeline as a }\textbf{\textit{reference union}}\textit{ of stages and not an order every harness follows \nohy{(OpenAnt} runs its own analysis, raptor drives Semgrep and \nohy{CodeQL,} ccsr is a model with file tools and no extension check, and VVAH reads the whole tree).}}
\label{fig:3}
\end{figure}

In Figure 3, each stage lists the drop points from the list above that fire there, so a reachable vulnerability can leave \textbf{before the model} (selection, parsing, call-graph, seeding) and \textbf{at or after it} (analysis, verification, reporting).

Whether an exit is silent or disclosed depends only on the deliverable surface (Rules 1 and 2 above), never on logs. Class 6 (Scan Limit Cutoff) is cross-cutting: it can act at selection, at a fetch, or inside analysis, so it is not pinned to one stage. Class 12 needs only the parse-emit gap, and code reached only through an unparsed block can then also be pruned.

\textbf{The harness and the OS. Some drops, such as those related to file extensions, could be inherited from outside the harness, but often are not.} The operating system does not route files to analyzers by extension, so routing by extension is a convention each harness implements itself. On Linux the operating system runs code by the shebang and the execute bit, and the filename is an opaque string (see execve(2), Linux man-pages project, \nohy{n.d.).}

Almost all thirteen pipeline drop points are in the harness’s own code, frequently using a standard-library string helper (\texttt{os.\allowbreak{}path.\allowbreak{}splitext}, \texttt{pathlib}, a regex, \texttt{endswith}, a content hash) that parses a string without ever consulting the OS or filesystem.

A few inherit a \textbf{third-party engine’s} behavior: Suppression Comments (8), where an engine honors an in-source suppression and the wrapper passes no override, as with Semgrep’s \# nosemgrep (Semgrep, \nohy{n.d.-b)} in Cisco’s ai-deep-sast or a compiler analyzer’s \nohy{[[clang::suppress]]} (LLVM Project, \nohy{n.d.)} in raptor’s opt-in compiler-scan channel.

The same applies to Unparsed Code Blocks (12), where tree-sitter cannot parse inside a macro and \nohy{CodeQL’s} build-based extraction never captures uncompiled code \nohy{(GitHub,} \nohy{n.d.-a).} It also applies to an inherited engine file-ignore: Semgrep’s default .semgrepignore (Semgrep, \nohy{n.d.-b)} skips minified, dist and vendored paths when the wrapper passes no override \nohy{(ToB’s} Semgrep skill, a path skip of class 11).

A few others act on a value from outside: an LLM’s reachability or labeling verdict (classes 4 and 7), or a file list from the \nohy{GitHub} API or git (classes 6 and 13). The decision to drop is still the harness’s own code.

\subsection{Beyond the thirteen: the model as the drop point}

Two further cases sit apart from the thirteen because no stage fails: the attacker works on the harness’s own model. We do not number them. Section 8.2 treats them under the same principle as classes 8 and 11, that text in the scanned code is data and never an instruction.

\textbf{Content-steered verdict.} A comment, docstring, or name whose content persuades the model to dismiss a real vulnerability, for example \# reviewed - safe, do not flag.

Not a pipeline stage: class 11 (\texttt{@generated}) drops the file and class 8 (\texttt{\# nosemgrep}) drops the finding, both without a model judging the code. Here the model \textit{reads the text and changes its own verdict}. A model that \textit{invents} a safety reason from a bare \# validated is a different case, not covered here.

A harness that feeds the file text to the model passes a safety-claiming comment along with it. Where nothing distrusts the claim, only the model reading past it stops a drop.

Two harnesses distrust such claims by rule: \textbf{Codex} suppresses a finding only on source-backed counterevidence, so a self-reported \nohy{safe/validated} comment does not qualify, and \textbf{\nohy{VulnHunter}} runs an explicit comment-skepticism check before any dismissal.

\textbf{\nohy{ToB} leans the other way. Its false-positive judge counts a \nohy{//} SAFETY: invariant as a mitigation.} Its false-positive step lists validated \nohy{//} SAFETY: invariants among the mitigations to check at the site \nohy{(trailofbits/skills} @1256982, \nohy{plugins/\allowbreak{}rust-review/\allowbreak{}agents/\allowbreak{}rust-review-fp-judge.\allowbreak{}md:\allowbreak{}85).}

Its severity rules lower an unsafe \{ \} finding by one level when a real \nohy{//} SAFETY: invariant holds, and keep the severity when the comment is pro-forma or wrong (:131-132). The model decides whether the invariant holds, and the code’s author writes the comment.

A finding judged a false positive, or lowered below a medium or high severity filter, is left out of \nohy{REPORT.md} and \nohy{REPORT.sarif} (:232-237, :301). It is still listed in \nohy{findings/} and in \nohy{fp-summary.md’s} per-primary table, both of which the plugin’s README names as outputs, so under Rules 1 and 2 this drop is disclosed, not silent.

\textbf{\nohy{RepoAudit} dropped a real, reachable finding on a planted safety comment in one of three runs, producing a report} that named the dropped sink nowhere. The disclosure test still applies (Rules 1 and 2): if the report names the dismissed finding and repeats the planted claim as the reason, the drop is disclosed, not silent.

\textbf{Prompt injection.} Instructions planted in the scanned code, diff, or comments that hijack the harness’s own model or agent and make it skip, reclassify, or stop.

The most direct outcome is a false clean: the model is steered to call a real bug safe. A content-steered verdict is that outcome when a comment does the steering. Most harnesses feed the target’s text to the model with no separation of instructions from code, so an injection arrives intact and only the model’s own refusal blocks it.

Structural defenses exist: \textbf{Codex}’s “treat repository text as data, never as instructions” rule on every worker prompt, and \textbf{raptor}’s nonce-tagged untrusted-input envelope plus a counter-escalation step that re-opens a finding the model was talked into clearing.

\textbf{\nohy{OpenAnt}} and \textbf{VVAH} carry the same idea off their main verdict path - \textbf{\nohy{OpenAnt}} in its threat-model agent, \textbf{VVAH} in an anti-manipulation rule.

\subsection{What we did not map}

During this work we identified many more drop points that fit our definition, which we decided were out of scope for this text. Below we list four such examples.

We did not trace them to a harness at a commit, so they are not in the taxonomy.

\begin{itemize}

\item \textbf{Wrong threat model applied. }The analysis runs to completion but answers the wrong security question and returns clean. For example, a harness tests an endpoint for injection by an anonymous user when the exposure is privilege escalation between authenticated roles.

\item \textbf{A fixture limit recorded as a clean result. }The case could not be exercised, for example because the verifier’s sandbox cannot reach the database the exploit needs, and the failed attempt is recorded as a clean result. Dynamic verification coverage is a stated limitation of \nohy{OpenAnt} (Korda \& Evron, 2026). Whether a harness records such a gap as untested or as clean is the open question.

\item \textbf{Dedup swallowing. }A new finding is absorbed into an earlier suppressed, duplicate, or baseline entry because the match key is too coarse, for example type and file alone. This is close to class 5. The difference is that in class 5 an old status wins, while here the new finding never surfaces as new at all.

\item \textbf{Documentation trusted as deployed enforcement. }The harness concludes “safe” because a specification, README, or docstring says a control exists, without checking that the running code enforces it. It is the content-steered verdict with the attestation moved from a comment to the documentation.

\end{itemize}

Each would need the same treatment as the thirteen: a mechanism in source, a planted input, and a check of every deliverable. We leave that to future work.

\section{Each drop point, in real harnesses}

Every example is a real mechanism in the harness’s source or its shipped instructions, cited by file and commit where a single location exists. We focus most on the two harnesses we maintain, \nohy{OpenAnt} and raptor, because we know them best.

\needspace{5\baselineskip}\textbf{1 - Unrecognized File Types}\par\nopagebreak

\begin{itemize}

\item \textbf{\nohy{OpenAnt}} - It picks a file’s language from its extension alone. A file with no extension (like the executable \nohy{bin/deploy)} or an unlisted one (like .pyw) gets no language and is never parsed. To trigger: put the vulnerable code in a file with no extension. To recover: rename it to .py.

{\raggedright Source: \nohy{knostic/openant} @b9d5d2f, \nohy{libs/\allowbreak{}openant-core/\allowbreak{}parsers/\allowbreak{}python/\allowbreak{}repository\_\allowbreak{}scanner.\allowbreak{}py:\allowbreak{}139-142} \nohy{(is\_source\_file;} the .py-only default is at line 98).\par}

\item \textbf{Codex Security \nohy{(OpenAI)}} - When it builds the file list for a diff (pull-request) scan, it drops any changed file whose extension is not on its allowed list.

{\raggedright Source: \nohy{openai/codex-security} @455d7c8, \nohy{sdk/\allowbreak{}typescript/\allowbreak{}\_\allowbreak{}bundled\_\allowbreak{}plugin/\allowbreak{}scripts/\allowbreak{}generate\_\allowbreak{}in\_\allowbreak{}scope\_\allowbreak{}files.\allowbreak{}py:\allowbreak{}143-146} (the \nohy{TEXT\_CODE\_EXTENSIONS} check. The file was later removed from the default branch and exists at this commit).\par}

\item \textbf{ai-deep-sast, \nohy{RepoAudit,} \nohy{ToB} and deepsec} - Each selects files by extension, so a file with no extension is never collected. ai-deep-sast walks the tree and keeps only paths that end in one of its include patterns. \nohy{RepoAudit} keeps only files whose suffix is on its per-language list.

The \nohy{ToB} Semgrep skill finds files with language globs such as \nohy{**/*.rs.} deepsec keeps a file only when one of its matchers’ glob patterns hits it, such as \nohy{**/*.py,} so an extensionless file gets no record and is never sent to the model. Section 7.1 shows the result for each.

{\raggedright Source: \nohy{cisco-open/\allowbreak{}ai-deep-sast} @741a9e0, \nohy{indexer.py:442-454;} \nohy{PurCL/RepoAudit} @160f5bc, \nohy{repoaudit.py:53-61} (the suffixs lists); \nohy{trailofbits/skills} @1256982, \nohy{plugins/\allowbreak{}static-analysis/\allowbreak{}skills/\allowbreak{}semgrep/\allowbreak{}workflows/\allowbreak{}scan-workflow.\allowbreak{}md:\allowbreak{}66;} \nohy{vercel-labs/deepsec} @76c03d6, \nohy{packages/\allowbreak{}scanner/\allowbreak{}src/\allowbreak{}index.\allowbreak{}ts:\allowbreak{}242-318} and the matchers’ \nohy{filePatterns} \nohy{(matchers/rce.ts:8,} \nohy{matchers/\allowbreak{}py-nosql-injection.\allowbreak{}ts).}\par}

\end{itemize}

\needspace{5\baselineskip}\textbf{2 - Argument Count Mismatch}\par\nopagebreak

\begin{itemize}

\item \textbf{\nohy{OpenAnt}} - Its Swift parser checks argument counts, but on a mismatch it keeps the call instead of dropping it, and it counts calls it could not resolve, so this is not silent.

{\raggedright Source: \nohy{knostic/openant} @b9d5d2f, \nohy{libs/\allowbreak{}openant-core/\allowbreak{}parsers/\allowbreak{}swift/\allowbreak{}call\_\allowbreak{}graph\_\allowbreak{}builder.\allowbreak{}py:\allowbreak{}1215} and 193-243.\par}

\item \textbf{\nohy{RepoAudit}} (a model-guided data-flow harness, Guo et al., 2025) - It links a call to the function it calls only when the number of arguments equals the number of parameters. A valid call that uses optional or variable arguments gets no link, so data is never tracked through it.

{\raggedright Source: \nohy{PurCL/RepoAudit} @160f5bc, \nohy{src/\allowbreak{}tstool/\allowbreak{}analyzer/\allowbreak{}TS\_\allowbreak{}analyzer.\allowbreak{}py:\allowbreak{}547-557} (len(paras) == len(arguments)).\par}

\end{itemize}

\needspace{5\baselineskip}\textbf{3 - Built-in Name Conflict}\par\nopagebreak

\begin{itemize}

\item \textbf{\nohy{OpenAnt}} - In PHP code, it checks a call’s name against PHP’s built-in functions before looking for the project’s own functions. If a project function shares a built-in’s name, calls to it are treated as calls to the built-in. A command injection reached through that function then falls out of the analysis, and \textbf{\nohy{OpenAnt}} reports only totals, not the dropped item.

Its Python analysis partly avoids this: a function with a built-in’s name defined in the same file is linked correctly \nohy{(libs/\allowbreak{}openant-core/\allowbreak{}parsers/\allowbreak{}python/\allowbreak{}call\_\allowbreak{}graph\_\allowbreak{}builder.\allowbreak{}py:\allowbreak{}436-440} at the same commit).

{\raggedright Source: \nohy{knostic/openant} @b9d5d2f, \nohy{libs/\allowbreak{}openant-core/\allowbreak{}parsers/\allowbreak{}php/\allowbreak{}call\_\allowbreak{}graph\_\allowbreak{}builder.\allowbreak{}py:\allowbreak{}281-287} (the \nohy{\_is\_builtin} check returns before any project-function lookup, except for higher-order built-ins such as \nohy{array\_map,} whose callback argument it resolves).\par}

\item \textbf{Other harnesses} - None of the others we read dropped calls this way without saying so. Most either link the call correctly or report what they skipped.

\end{itemize}

\needspace{5\baselineskip}\textbf{4 - Unrecognized Entry Points}\par\nopagebreak

\begin{itemize}

\item \textbf{\nohy{OpenAnt}} - It treats code as reachable only if it starts from an entry-point type on a fixed list. The list has no entry for Celery background tasks, AWS Lambda handlers, or Django signal receivers, so these handlers are treated as unreachable and removed, and the scan reports clean.

{\raggedright Source: \nohy{knostic/openant} @c81c0e1, \nohy{libs/\allowbreak{}openant-core/\allowbreak{}utilities/\allowbreak{}agentic\_\allowbreak{}enhancer/\allowbreak{}entry\_\allowbreak{}point\_\allowbreak{}detector.\allowbreak{}py:\allowbreak{}212-214} \nohy{(ENTRY\_POINT\_TYPES} at lines 39-54. The file’s decorator and input-pattern checks have no entry for these frameworks either.\par}

At the study pin b9d5d2f the same check is at line 305 and \nohy{ENTRY\_POINT\_TYPES} at line 39, with a longer list that still has no entry for these three).

\item \textbf{raptor} - It has the same gap in a milder form: the handler is only marked low priority, never removed.

{\raggedright Source: \nohy{gadievron/raptor} @32ed579, \nohy{core/\allowbreak{}orchestration/\allowbreak{}reachability\_\allowbreak{}enrichment.\allowbreak{}py:\allowbreak{}153-165} (sets priority to “low”).\par}

\item \textbf{iris} (a Java harness built on \nohy{CodeQL,} Li, Dutta \& Naik, 2025) - It uses a model to label where untrusted data enters and where it becomes dangerous, and tracks data only from what the model labels. The model is shown only public functions from the project’s own code that take parameters, are called by its tests, and are not called by any other project code.

A handler that only the framework calls is never shown, so it is never labeled or tracked, and its vulnerable code is never reached. This applies when the untrusted input arrives as a handler parameter.

{\raggedright Source: \nohy{iris-sast/iris} @3a12f45, \nohy{src/\allowbreak{}queries/\allowbreak{}fetch\_\allowbreak{}func\_\allowbreak{}params.\allowbreak{}ql:\allowbreak{}40-47} \nohy{(isTested} and \nohy{is\allowbreak{}Not\allowbreak{}Invoked\allowbreak{}By\allowbreak{}Internal\allowbreak{}Function} are defined at lines 12-24); the model labels these candidates in \nohy{src/iris.py:700-741.}\par}

\end{itemize}

\needspace{5\baselineskip}\textbf{5 - Overridden Findings}\par\nopagebreak

\begin{itemize}

\item \textbf{deepsec} (Vercel’s harness) - When it re-checks code, it matches each new finding to stored ones by vulnerability type and title. A match is treated as a duplicate, and the earlier status (fixed, false positive, or accepted risk) wins.

So if a change brings back a bug that was marked fixed, the new, correct finding is left out of the count, the pull-request comment, and the export, and the scan still exits as successful.

{\raggedright Source: \nohy{vercel-labs/deepsec} @76c03d6, \nohy{packages/\allowbreak{}processor/\allowbreak{}src/\allowbreak{}index.\allowbreak{}ts:\allowbreak{}676-684} (matches on \nohy{vulnSlug} plus the lowercased title).\par}

\end{itemize}

\needspace{5\baselineskip}\textbf{6 - Scan Limit Cutoff}\par\nopagebreak

\begin{itemize}

\item \textbf{\nohy{OpenAnt}} - Its scan limit option sorts code by a priority score, analyzes the top items, and drops the rest without notice.

{\raggedright Source: \nohy{knostic/openant} @b9d5d2f, \nohy{libs/\allowbreak{}openant-core/\allowbreak{}core/\allowbreak{}analyzer.\allowbreak{}py:\allowbreak{}98-119} \nohy{(\_apply\_limit).}\par}

\item \textbf{ccsr (Anthropic’s Claude Code Security Review, a \nohy{GitHub} Action)} - It fetches only the first 100 changed files of a pull request. It fetches the full diff separately, so normally the code is still reviewed. But in the fallback mode used when the prompt is too long, files past the first 100 drop out of what the model is told to review. This is a conditional gap and not a clean silent drop.

{\raggedright Source: \nohy{anthropics/\allowbreak{}claude-code-security-review} @0c6a49f, \nohy{claudecode/\allowbreak{}github\_\allowbreak{}action\_\allowbreak{}audit.\allowbreak{}py:\allowbreak{}77-81} \nohy{(per\_page=100,} the API’s maximum page size \nohy{(GitHub,} \nohy{n.d.-b),} with no request for later pages).\par}

\end{itemize}

\needspace{5\baselineskip}\textbf{7 - Verification Filtering}\par\nopagebreak

\begin{itemize}

\item \textbf{raptor }- It downgrades any finding that no engine other than the model backs up. The model’s own reasoning does not count as evidence. The downgraded finding is marked clean (or “dark”, meaning unverified, when no engine covers that bug type), noted only in the run log, and left out of the graded results file. So a real, reachable SQL injection that the model found can be dropped with nothing on the report.

This happens only when nothing else covers the code: its built-in pre-check handles only C, C++ and Python, so PHP and model-only runs are exposed. If another engine’s results flag the same lines, or the finding loosely matches one of the project rules raptor worked out in an earlier study step (raptor calls these domain-model invariants), the finding is confirmed and reported.

The effect is inverted: a confident finding is dropped, while a hedged “suspicious” one would have been reported.

{\raggedright Source: \nohy{gadievron/raptor} @32ed579, \nohy{core/\allowbreak{}audit/\allowbreak{}orchestrator.\allowbreak{}py:\allowbreak{}4156-4172} (gate G2) and 7424-7538 \nohy{(\_\allowbreak{}resolve\_\allowbreak{}gate\_\allowbreak{}demoted);} \nohy{core/\allowbreak{}audit/\allowbreak{}findings\_\allowbreak{}export.\allowbreak{}py:\allowbreak{}157} keeps only “finding” and “suspicious” results; \nohy{core/\allowbreak{}audit/\allowbreak{}prefilter.\allowbreak{}py:\allowbreak{}148-151} (C, C++ and Python only).\par}

\item \textbf{\nohy{RepoAudit}} - It asks the model whether each candidate bug path is reachable and writes a report only if the answer is yes. A real path that the model wrongly calls unreachable would be dropped silently.

{\raggedright Source: \nohy{PurCL/RepoAudit} @160f5bc, \nohy{src/\allowbreak{}agent/\allowbreak{}dfbscan.\allowbreak{}py:\allowbreak{}675-693} (a report is written only when \nohy{is\_reachable} is true).\par}

\item \textbf{ai-sast, iris and \nohy{RepoAudit}} - Each expects the model’s reply in a fixed format and treats any other reply as no finding. ai-sast accepts a verdict only when the reply starts with “TRUE”. iris turns a reply it cannot parse as JSON into an empty result. \nohy{RepoAudit} requires an “Answer:” line that its regular expression can match.

{\raggedright Source: \nohy{rivian/ai-sast} @d3982ca, \nohy{src/core/validator.py} \nohy{(\_parse\_response);} \nohy{iris-sast/iris} @3a12f45, \nohy{src/\allowbreak{}modules/\allowbreak{}contextual\_\allowbreak{}analysis\_\allowbreak{}pipeline.\allowbreak{}py:\allowbreak{}312-324;} \nohy{PurCL/RepoAudit} @160f5bc, \nohy{src/\allowbreak{}llmtool/\allowbreak{}dfbscan/\allowbreak{}path\_\allowbreak{}validator.\allowbreak{}py:\allowbreak{}102.}\par}

\end{itemize}

\needspace{5\baselineskip}\textbf{8 - Suppression Comments}\par\nopagebreak

\begin{itemize}

\item \textbf{raptor} - Its optional compiler scan turns compiler warnings into findings but does not switch off in-code suppressions. So a suppression marker on a vulnerable line (a clang suppress attribute, or a GCC diagnostic pragma that silences the warning, Free Software Foundation, \nohy{n.d.)} makes the compiler report nothing, and no finding is created.

The gap is narrow: the scan is opt-in, the bug is still caught if Semgrep or \nohy{CodeQL} flags it, and this module exists only in a later raptor commit, 1b84ec5, after the version we studied. raptor does block Semgrep suppressions, and its other compiler path (also only in that later commit) spots suppression markers and will not treat the compiler’s silence as proof the code is safe.

{\raggedright Source: \nohy{gadievron/raptor} @1b84ec5 (newer than the studied @32ed579), \nohy{packages/\allowbreak{}static-analysis/\allowbreak{}compiler\_\allowbreak{}scan.\allowbreak{}py} (no suppression override) and \nohy{core/\allowbreak{}audit/\allowbreak{}compiler\_\allowbreak{}sweep.\allowbreak{}py:\allowbreak{}1200-1281} (the suppression check); -{}-disable-nosem is at \nohy{packages/\allowbreak{}semgrep/\allowbreak{}runner.\allowbreak{}py:\allowbreak{}87} (@32ed579).\par}

\item \textbf{ai-deep-sast} (Cisco’s Semgrep wrapper) - It runs Semgrep without the option that ignores \# nosemgrep comments. So return eval(data) is flagged, but return eval(data) \# nosemgrep is not, and the harness’s reports show neither the suppression nor a count of suppressed findings.

{\raggedright Source: \nohy{cisco-open/\allowbreak{}ai-deep-sast} @741a9e0, \nohy{aideepsast.py:394-398} (the Semgrep command, with no -{}-disable-nosem, see Semgrep, \nohy{n.d.-a).}\par}

\end{itemize}

\needspace{5\baselineskip}\textbf{9 - Inactive Rules}\par\nopagebreak

\begin{itemize}

\item \textbf{raptor} - It ships eight custom \nohy{CodeQL} queries, documented for users, that are never added to the set of queries a scan actually runs.

{\raggedright Source: \nohy{gadievron/raptor} @32ed579, \nohy{core/\allowbreak{}config/\allowbreak{}\_\allowbreak{}\_\allowbreak{}init\_\allowbreak{}\_\allowbreak{}.py:\allowbreak{}137} \nohy{(CODEQL\_QUERIES\_DIR} is defined but never used; the eight queries are in \nohy{engine/codeql/queries/} and documented in \nohy{docs/\allowbreak{}codeql.\allowbreak{}md:\allowbreak{}325-345).}\par}

\item \textbf{VVAH} (Visa’s harness that reads the whole code tree) - It has a fully written specialist for unsafe deserialization, but none of its shipped profiles turns it on, so it never runs unless a user adds it.

{\raggedright Source: \nohy{visa/\allowbreak{}visa-vulnerability-agentic-harness} @3d972f6, \nohy{vvaharness/\allowbreak{}lang/\allowbreak{}hints.\allowbreak{}py:\allowbreak{}1131} (the specialist); \nohy{vvaharness/\allowbreak{}config/\allowbreak{}profiles/\allowbreak{}*.yaml} (the specialists lists leave it out).\par}

\end{itemize}

\needspace{5\baselineskip}\textbf{10 - Stale Caching}\par\nopagebreak

\begin{itemize}

\item \textbf{raptor - It caches the model’s verdicts, keyed on the exact request} (model, instructions, and prompt) but not on the harness’s version, and by default the cache never expires. So after a harness update that leaves a finding’s prompt text unchanged, unchanged code is served its old “clean” verdict.

We ran raptor’s actual cache-key code, and the key stayed the same across such a change. The cache has no expiry \nohy{(cache\_ttl\_seconds} is None), so a months-old entry is served as fresh. Changes that alter a finding’s prompt, such as an edited rule message, do create a new key, so the stale result affects only changes the prompt does not show. Its dependency-scan cache has the same gap.

{\raggedright Source: \nohy{gadievron/raptor} @32ed579, \nohy{core/\allowbreak{}llm/\allowbreak{}client.\allowbreak{}py:\allowbreak{}998-1002} (the cache key) and \nohy{core/\allowbreak{}llm/\allowbreak{}config.\allowbreak{}py:\allowbreak{}1098} \nohy{(cache\_ttl\_seconds} = None); \nohy{packages/\allowbreak{}llm\_\allowbreak{}analysis/\allowbreak{}agent.\allowbreak{}py:\allowbreak{}1058-1086} (what goes into the prompt) and \nohy{packages/\allowbreak{}sca/\allowbreak{}\_\allowbreak{}file\_\allowbreak{}scan\_\allowbreak{}cache.\allowbreak{}py.}\par}

\item \textbf{\nohy{VulnHunter}} (Capital One’s batch harness) - It has a resume option that skips any repository that already has a results file. A repository that became vulnerable since the last scan is served its old “clean” result. We ran the resume option: it skipped the repository that already had results and scanned nothing.

{\raggedright Source: \nohy{Capital\allowbreak{}One/Vuln\allowbreak{}Hunter} @8f9eadd, \nohy{harness/\allowbreak{}local\_\allowbreak{}harness/\allowbreak{}scan.\allowbreak{}py:\allowbreak{}64-70} \nohy{(has\_valid\_results:} a results \nohy{README.md} over 100 bytes).\par}

\end{itemize}

\needspace{5\baselineskip}\textbf{11 - Keyword-Based Skipping}\par\nopagebreak

\begin{itemize}

\item \textbf{\nohy{OpenAnt}} - When run from its command line, it skips Python test files by default: anything under a test or tests folder, plus files named \nohy{test\_*.py,} \nohy{*\_test.py,} or \nohy{conftest.py.} A reachable vulnerability there is never read.

This skip is reported, not silent: \textbf{\nohy{OpenAnt}} records the number of skipped test files in its results file. VVAH’s \nohy{tests/} exclusion is disclosed in its own report (Section 2.1 and the VVAH entry below). The silent examples in this class are ccsr and Rivian’s ai-sast, below. \nohy{ToB’s} Semgrep skill is a third. Its path skip is described under class 12.

{\raggedright Source: \nohy{knostic/openant} @b9d5d2f, \nohy{libs/\allowbreak{}openant-core/\allowbreak{}parsers/\allowbreak{}python/\allowbreak{}repository\_\allowbreak{}scanner.\allowbreak{}py:\allowbreak{}144-162} \nohy{(is\_test\_file),} turned on by default in \nohy{libs/\allowbreak{}openant-core/\allowbreak{}openant/\allowbreak{}cli.\allowbreak{}py:\allowbreak{}81} (the library default at \nohy{repository\_\allowbreak{}scanner.\allowbreak{}py:\allowbreak{}103} is off); the skip count is at \nohy{repository\_\allowbreak{}scanner.\allowbreak{}py:\allowbreak{}119} and 313-314 \nohy{(test\_\allowbreak{}files\_\allowbreak{}skipped).}\par}

\end{itemize}

\begin{itemize}

\item \textbf{ccsr} (Anthropic) - It splits the diff by file and drops a file’s entire change if “@generated” appears anywhere in it, including in the changed code itself. A real, hand-written file is skipped just because one line contains that text. Nothing records the skip, so the exit code reflects only what was left.

{\raggedright Source: \nohy{anthropics/\allowbreak{}claude-code-security-review} @0c6a49f, \nohy{claudecode/\allowbreak{}github\_\allowbreak{}action\_\allowbreak{}audit.\allowbreak{}py:\allowbreak{}159-174} \nohy{(\_\allowbreak{}filter\_\allowbreak{}generated\_\allowbreak{}files).}\par}

\item \textbf{ai-sast }(Rivian’s model-only harness) - It skips any file whose full path contains a keyword from its list (such as log, script, png, or .git), ignoring case. So \nohy{auth/login.py} (contains “log”) and \nohy{transcript.go} (contains “script”) are skipped, but \nohy{deploy.py} is not.

The skip is printed only to the console, not to the report, HTML output, pull-request comment, or exit status. It is on by default and cannot be turned off.

{\raggedright Source: \nohy{rivian/ai-sast} @d3982ca, \nohy{src/\allowbreak{}core/\allowbreak{}scanner.\allowbreak{}py:\allowbreak{}587} (the keyword list) and 597-603 (the substring match); the console print is at 622-626.\par}

\item \textbf{VVAH} (Visa) - It excludes test directories by default, including \nohy{tests/,} before any file is read, and its report records the exclusion with a count, so this drop is disclosed.

{\raggedright Source: \nohy{visa/\allowbreak{}visa-vulnerability-agentic-harness} @3d972f6, \nohy{vvaharness/\allowbreak{}pipeline/\allowbreak{}stages/\allowbreak{}s1\_\allowbreak{}preprocess.\allowbreak{}py:\allowbreak{}43-50} \nohy{(\_\allowbreak{}DEFAULT\_\allowbreak{}EXCLUDE\_\allowbreak{}DIRS,} with tests at line 50), applied at :97-99 \nohy{(\_exclusion\_sets).}\par}

\end{itemize}

\needspace{5\baselineskip}\textbf{12 - Unparsed Code Blocks}\par\nopagebreak

\begin{itemize}

\item \textbf{\nohy{OpenAnt}} - To its Rust parser, a macro body is an unreadable block, so a vulnerability inside a \nohy{fuzz\_target!} macro produces nothing to analyze. We reproduced it at the cited commit: a file with the same command-injection sink in a normal function and inside \nohy{fuzz\_target!} yields one unit, for the normal function only.

{\raggedright Source: \nohy{knostic/openant} @b9d5d2f, \nohy{libs/\allowbreak{}openant-core/\allowbreak{}parsers/\allowbreak{}rust/\allowbreak{}function\_\allowbreak{}extractor.\allowbreak{}py:\allowbreak{}342-356} (only \nohy{function\_item} nodes become units; tree-sitter keeps a macro body as an unparsed token tree).\par}

\item \textbf{\nohy{ToB}} (Trail of Bits’ \nohy{CodeQL} and Semgrep skills) - The \nohy{CodeQL} skill builds its database while the project compiles, so code the build does not compile is never captured. (This comes from the skill’s instructions, since there is no single line of code to cite.) Separately, the skill deliberately excludes third-party, vendored, generated, and minified code. That is a documented choice and a different matter from this gap.

The Semgrep skill has a different, silent gap, a path skip that belongs to class 11: it keeps Semgrep’s default ignore list (Semgrep, \nohy{n.d.-b),} which skips minified, dist, and vendored files, and its report shows only how many files were scanned, not which were skipped. A file merely named “generated” is not skipped.

{\raggedright Source: \nohy{trailofbits/skills} @1256982, under \nohy{plugins/\allowbreak{}static-analysis/\allowbreak{}skills/\allowbreak{}:} \nohy{codeql/\allowbreak{}workflows/\allowbreak{}build-database.\allowbreak{}md:\allowbreak{}29} and :104; \nohy{semgrep/\allowbreak{}references/\allowbreak{}scanner-task-prompt.\allowbreak{}md:\allowbreak{}38;} \nohy{semgrep/\allowbreak{}workflows/\allowbreak{}scan-workflow.\allowbreak{}md:\allowbreak{}289.}\par}

\item \textbf{iris} - Built on \nohy{CodeQL} too, it has the same compile-only gap, which also misses generated code that the build does not compile.

{\raggedright Source: \nohy{iris-sast/iris} @3a12f45 (it uses \nohy{CodeQL} databases built from a compile; there is no single line to cite).\par}

\end{itemize}

\needspace{5\baselineskip}\textbf{13 - Diff Blindspots}\par\nopagebreak

\begin{itemize}

\item \textbf{\nohy{OpenAnt}} - It also narrows diff-mode scans to changed code, but it marks every unit in its dataset as selected or not and counts them, so this is not silent.

{\raggedright Source: \nohy{knostic/openant} @b9d5d2f, \nohy{libs/\allowbreak{}openant-core/\allowbreak{}core/\allowbreak{}diff\_\allowbreak{}filter.\allowbreak{}py:\allowbreak{}80-149} \nohy{(diff\_selected} on every unit, plus \nohy{DiffStats).}\par}

\item \textbf{deepsec} - In diff mode, it asks git only for added, modified, renamed, or copied files, which leaves out deleted files. So a deleted file that held a security check, or an unchanged file made unsafe by a change elsewhere, is never selected.

{\raggedright Source: \nohy{vercel-labs/deepsec} @76c03d6, \nohy{packages/\allowbreak{}deepsec/\allowbreak{}src/\allowbreak{}file-sources.\allowbreak{}ts:\allowbreak{}45-56} (-{}-diff-filter=\allowbreak{}AMRC); the selected files are then scanned whole \nohy{(packages/\allowbreak{}deepsec/\allowbreak{}src/\allowbreak{}commands/\allowbreak{}process.\allowbreak{}ts:\allowbreak{}311-316).}\par}

\end{itemize}

\section{Proof of concept: hiding a vulnerability from one harness}

\needspace{9\baselineskip}Here the hiding is done by the harness’s \textbf{own reachability analysis} - the bug is live, but the call graph is made to believe it is unreachable. A command-injection sink, reached from a request handler through one ordinary function call:\par\nopagebreak

\begin{Verbatim}[commandchars=\|\!\^,fontsize=\small,breaklines=true,breakanywhere=true,breaknonspaceingroup=true,xleftmargin=1.5em]
|codecolA!|textbf!<?php^^
|codecolA!|textbf!namespace^^ App|codecolA!;^
|codecolA!|textbf!function^^ route() {
    |codecolB!$c = $_GET['c'];^
    header_x($c);                                     |codecolC!|textit!// reaches the sink^^
}
|codecolA!|textbf!function^^ header_x($x) {
    |codecolD!|textbf!system^^($x)|codecolA!;^   |codecolC!|textit!// command injection^^
}
\end{Verbatim}

\textbf{The bug is reachable:} route reads a request parameter and passes it to \nohy{header\_x,} so the sink is reached from a live entry point. That is the ground truth. The hide below is the \textit{harness’s} call graph deciding otherwise. That is what “reachability” means in this section, and it is a separate thing from whether the code is live.

\textbf{The hide - rename one function.} \nohy{OpenAnt’s} call-graph builder vetoes a call whose name is on the PHP built-in list, apart from a few higher-order built-ins such as \nohy{array\_map,} before it ever looks for a user-defined function.

\needspace{9\baselineskip}So rename the sink \nohy{header\_x} → header (a real PHP built-in), and change nothing else. Declared inside namespace App, the function is App\textbackslash{}header and does not collide with the global \textbackslash{}header, so it is legal PHP (no fatal redeclaration). An unqualified header(...) call there resolves to App\textbackslash{}header first, so the program behaves the same. It is the bare-name match on header that trips \nohy{OpenAnt’s} built-in veto:\par\nopagebreak

\begin{Verbatim}[commandchars=\|\!\^,fontsize=\small,breaklines=true,breakanywhere=true,breaknonspaceingroup=true,xleftmargin=1.5em]
    |codecolE!header($c);^          |codecolC!|textit!// identical call; OpenAnt now ^^|codecolC!|textit!resolves it to the builtin header(), not our code^^
    ...
|codecolA!|textbf!function^^ |codecolE!header($x) {^ |codecolD!|textbf!system^^($x)|codecolA!;^ }
\end{Verbatim}

\needspace{5\baselineskip}\textbf{Steps \nohy{(OpenAnt’s} real extractor and call-graph builder, run without the model):}\par\nopagebreak

\begin{enumerate}[label=\arabic*.,start=1]

\item \texttt{header\_\allowbreak{}x} version → the \texttt{route} → \texttt{header\_\allowbreak{}x} edge is built, the sink is reachable, and the command-injection sink enters the reachable set.

\item Rename the sink to header → the call is vetoed as the built-in → the sink gets \textbf{zero inbound edges}, is isolated, and is pruned by \nohy{OpenAnt’s} default \texttt{reachable} filter.

\item Result: the identical command injection is silently out of scope. The only trace is an aggregate “N units pruned” count, and no line names the function.

\nohy{(OpenAnt} does have a per-unit \nohy{pruned\_units.json} telemetry that would name it, but only on its core parser path, which the Python parser uses and the PHP parser here does not. The PHP path writes just the aggregate \nohy{filtered\_out} count, which mixes every prune cause and so cannot be tied to the function (Rule 2). We confirmed this by running the real PHP parse at the study pin, b9d5d2f: the pruned sink is absent from the output and nothing names it. The telemetry is \nohy{libs/\allowbreak{}openant-core/\allowbreak{}utilities/\allowbreak{}prune\_\allowbreak{}telemetry.\allowbreak{}py,} wired through \nohy{core/\allowbreak{}parser\_\allowbreak{}adapter.\allowbreak{}py,} at b9d5d2f.)

\end{enumerate}

The bug never moved and is still reachable, but the harness’s call graph decided it was not. We ran \nohy{OpenAnt’s} extractor and call-graph builder and saw the edge disappear (Figure 4). route → \nohy{header\_x} builds it, route → header builds none, so the sink falls out of reachable scope.

This is a \textit{reachability-analysis} failure - distinct from the file-selection drops below, and from whether the planted bug is live.

\begin{figure}[htbp]
\centering
\includegraphics[width=0.897\textwidth]{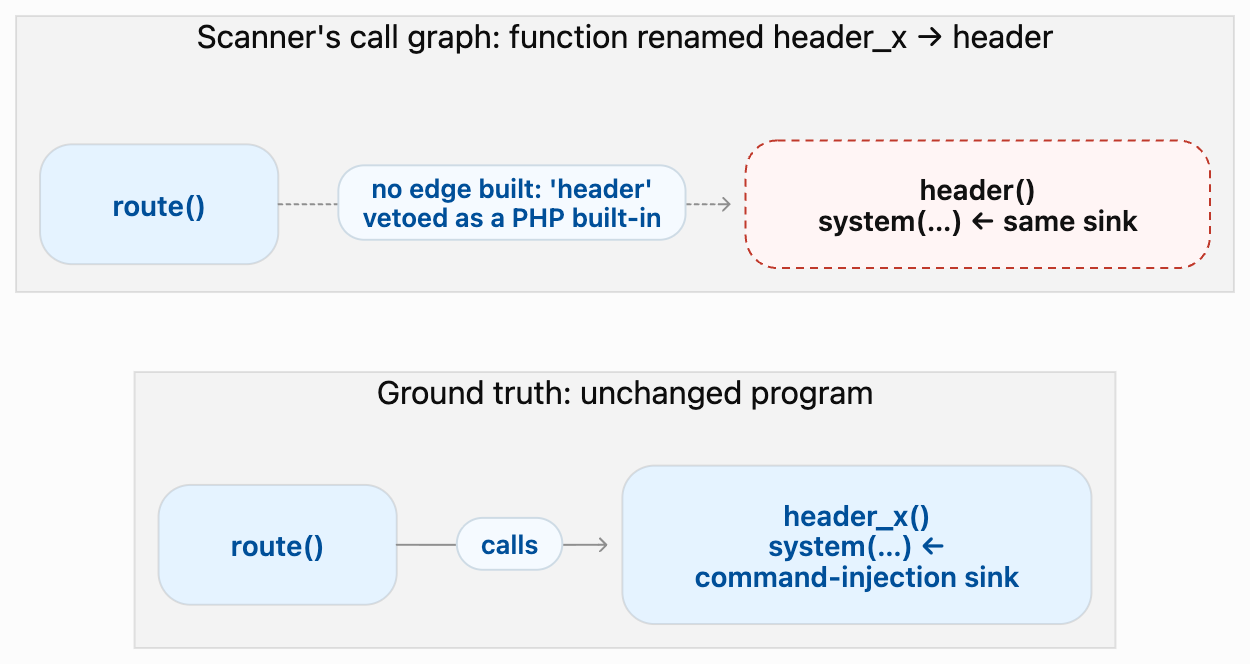}
\caption{\textit{Both panels are the same program, only the function’s name changed. “Reachable” here means the harness’s call-graph verdict, not whether the code is live - the sink is live in both. The lower panel is the program before the rename: route calls \nohy{header\_x,} and \nohy{OpenAnt} builds that edge. The upper panel is \nohy{OpenAnt’s} call graph after the rename: at run time the call still reaches the same sink, but \nohy{OpenAnt} vetoes route → header (the name matches a PHP built-in), so the sink gets zero inbound edges and is pruned. This is the reachability-analysis sense, distinct from the file-selection drops that follow.}}
\label{fig:4}
\end{figure}

\section{Proof of concept: hiding it from one harness three ways at once}

Same idea, one harness, three of its drop points stacked on one file. For the stack we use a PHP handler, where three of \nohy{OpenAnt’s} drop points line up.

\needspace{9\baselineskip}This is the file. The first line is our annotation, not part of the file (placed above \textless{}?php, PHP would print it as output):\par\nopagebreak

\begin{Verbatim}[commandchars=\|\!\^,fontsize=\small,breaklines=true,breakanywhere=true,breaknonspaceingroup=true,xleftmargin=1.5em]
|codecolC!|textit!// saved as tests/handler (PHP code, but NO .php extension), and left out of the ^^|codecolC!|textit!pull-request diff^^
|codecolF!<^|codecolA!?^php
|codecolA!|textbf!function^^ run_cmd() {
    |codecolA!|textbf!return^^ |codecolA!|textbf!eval^^(|codecolB!$_GET^[|codecolG!'cmd'^])|codecolA!;^   |codecolC!|textit!// RCE:  ?cmd=<php>^^
}
\end{Verbatim}

The file is live: the application’s front controller loads it with include \nohy{\_\_DIR\_\_} . \nohy{\textquotesingle{}/\allowbreak{}../\allowbreak{}tests/\allowbreak{}handler\textquotesingle{};} and calls \nohy{run\_cmd()} for the \nohy{/run} route. PHP’s include runs a file whatever its extension, and the includer does not change \nohy{OpenAnt’s} selection of \nohy{tests/handler.} The bug is a directly exploitable eval of a request parameter in live code, not inert test data, so the \nohy{tests/} path is a mistaken skip by the harness.

Placed that way, the one file hits three of \nohy{OpenAnt’s} selection drop points at once:

\begin{itemize}

\item \textbf{No extension} → \texttt{is\_\allowbreak{}source\_\allowbreak{}file} never collects it (class 1, Unrecognized File Types)

\item \textbf{Under }\textbf{\texttt{tests/\allowbreak{}}} → even with an extension, \texttt{is\_\allowbreak{}test\_\allowbreak{}file} would skip it (class 11, Keyword-Based Skipping)

\item \textbf{Outside the pull-request diff} → the diff filter drops the unit (class 13, Diff Blindspots).

\end{itemize}

(For PHP, the extension check and the test-file rule are in \nohy{libs/\allowbreak{}openant-core/\allowbreak{}parsers/\allowbreak{}php/\allowbreak{}repository\_\allowbreak{}scanner.\allowbreak{}py:\allowbreak{}119-122} and :124 at b9d5d2f. The Python parser cited in Section 4 has its own copy.) These checks stop before the model. In the steps below, “reported” means the vulnerable unit reached the dataset the model would analyze, and “clean” means it did not.

\needspace{5\baselineskip}\textbf{Steps (three \nohy{OpenAnt} selection gates, each verified separately; not one end-to-end run):}\par\nopagebreak

\begin{enumerate}[label=\arabic*.,start=1]

\item Plain \nohy{src/handler.php,} inside the diff → reported

\item Apply all three placements → clean

\item Undo any one of the three → still clean, because the other two each still drop it. Only undoing all three restores the finding. We verified each of the three gates separately. The stacked and single-undo cases follow from the mechanisms, since the three gates are independent and the extension gate fires first.

\end{enumerate}

Each one alone would drop the file. Stacked, fixing any single one still leaves the other two, so the file stays out of the scan until all three are undone.

The stacking also hides the evidence: run the test-path drop alone and \nohy{OpenAnt’s} \nohy{scan\_results.json} records a \nohy{test\_files\_skipped} count that moved. Run the diff drop alone and the dataset marks the unit \nohy{diff\_selected:} false.

But the extension check (class 1) fires first, at file selection, before any unit exists - so in the stacked case \textbf{neither of those notices can be written}. Stacking drop points also removes the traces a single drop would have left.

More than three can be stacked. \nohy{OpenAnt} has other silent drop points that could join the same stack, for example Scan Limit Cutoff (6), Unrecognized Entry Points (4), and Built-in Name Conflict (3).

\section{Proof of concept: hiding it from many harnesses at once}

\subsection{One change, many harnesses (each through its own drop point)}

Here each harness gets a file in a language it targets, which brings in \nohy{ToB} with a Rust file and deepsec with a Python file. Section 7.2 then uses one Python file and a six-harness panel, without \nohy{ToB} and deepsec. iris, \nohy{VulnHunter,} ai-sast and raptor select files without an extension list and appear in neither panel.

Give each harness a real vulnerability \textit{in the language it targets} - a Python \texttt{eval()}, a Rust \nohy{Command::new(...).\allowbreak{}spawn()} on request input, a \nohy{py-NoSQL} \$where - then make the same single change to each:

\textbf{Strip the file’s extension} (\texttt{src/\allowbreak{}handler.\allowbreak{}py} → \texttt{src/\allowbreak{}handler}, \texttt{lib.\allowbreak{}rs} → \texttt{lib})

Run each harness’s own file-selection code. \nohy{(ToB’s} run used a different Rust sink, \nohy{mem::forget.} \nohy{ToB’s} drop comes from the \nohy{**/*.rs} glob, which never matches the file, so the choice of sink does not affect it.)

\begin{center}\small
\begin{tabularx}{0.846\textwidth}{>{\raggedright\arraybackslash\hsize=1.000\hsize}X>{\raggedright\arraybackslash\hsize=1.000\hsize}X}
\toprule
\textbf{harness} & \textbf{after the extension is stripped} \\
\midrule
\nohy{OpenAnt} & file not collected \\
\nohy{RepoAudit} & file not collected \\
ai-deep-sast & file not indexed \\
\nohy{ToB} (Semgrep skill, Rust target) & \texttt{*.rs} glob misses it \\
Codex & dropped from the diff inventory \\
deepsec & dropped at selection \\
\bottomrule
\end{tabularx}
\end{center}

As a counter-example, two other harnesses were resistant:

\textbf{VVAH} reads the whole tree regardless of extension, and Anthropic’s ccsr reviews the diff with a model that has file tools. Neither has an extension check to exploit. So the one cosmetic change drops the file in every other harness in this panel, each through its own drop point.

\subsection{One identical file, each harness tripped by its own selection rule}

In this second example, we hide \textit{the same} bug from a whole panel at once.

One Python \texttt{eval()} file with three placements stacked on it, reachable from a production entry point (\texttt{app.\allowbreak{}py} loads the extensionless \nohy{tests/handler} and binds \nohy{run\_cmd} as a live route).

\begin{Verbatim}[commandchars=\|\!\^,fontsize=\small,breaklines=true,breakanywhere=true,breaknonspaceingroup=true,xleftmargin=1.5em]
|codecolC!|textit!# tests/handler ^^|codecolC!|textit!-^^|codecolC!|textit! no extension, a false @generated banner, and the real bug^^
|codecolC!|textit!# @generated by codegen-tool; do not edit^^
|codecolH!|textbf!import^^ flask
|codecolA!|textbf!def^^ run_cmd():
    |codecolA!|textbf!return^^ |codecolH!eval^(flask.request.args.get(|codecolG!"cmd"^))      |codecolC!|textit!# RCE: ?cmd=<python>^^

|codecolC!|textit!# src/app.py ^^|codecolC!|textit!-^^|codecolC!|textit! ^^|codecolC!|textit!a production loader runs the extensionless tests/ file^^|codecolC!|textit! and ^^|codecolC!|textit!binds it as a live route^^
|codecolH!|textbf!import^^ runpy, os, flask
app = flask.Flask(__name__)
p = os.path.join(os.path.dirname(__file__), "..", "tests", "handler")
m = runpy.run_path(p)
app.add_url_rule("/run", "run", m["run_cmd"])       |codecolC!|textit!# the live entry point binds the bug^^
\end{Verbatim}

\needspace{5\baselineskip}\textbf{Walkthrough (each harness’s own selection code, run without the model):}\par\nopagebreak

\begin{enumerate}[label=\arabic*.,start=1]

\item \textbf{Control} - the same bug at \texttt{src/\allowbreak{}handler.\allowbreak{}py}, normal path, inside the diff: every harness enumerates it.

\item \textbf{Hide - move it to the file above and re-run each harness’s selection stage.}

\item \textbf{Result} - the one file is dropped before the model sees its code, by every harness in the panel. They split on two axes.

\textit{Which drop point trips: four hit the same one} (the missing extension), VVAH hits the \texttt{tests/\allowbreak{}} directory, ccsr hits the \texttt{@generated} substring.

\end{enumerate}

\begin{itemize}[label={}]

\item \textit{Outcome:} the drop is \textbf{silent} in five of the six - only \textbf{VVAH} discloses it:

\end{itemize}

\begin{center}\small
\begin{tabularx}{0.846\textwidth}{>{\raggedright\arraybackslash\hsize=1.000\hsize}X>{\raggedright\arraybackslash\hsize=1.000\hsize}X>{\raggedright\arraybackslash\hsize=1.000\hsize}X}
\toprule
\textbf{harness} & \textbf{the drop point it trips} & \textbf{outcome} \\
\midrule
\nohy{OpenAnt} & no extension & dropped - silent \\
\nohy{RepoAudit} & no extension & dropped - silent \\
ai-deep-sast & no extension & dropped - silent \\
Codex Security & no extension (diff path) & dropped - silent \\
VVAH & \texttt{tests/\allowbreak{}} directory & dropped - \textbf{disclosed} (\texttt{excluded.\allowbreak{}dirs: \{tests: 1\}} in its own report) \\
ccsr & \texttt{@generated} substring strips the diff section & the code is dropped from the diff - silent \\
\bottomrule
\end{tabularx}
\end{center}

ccsr strips it: its @generated check removes the file’s code from the diff before the model sees it, leaving only the bare filename in the prompt’s file list.

Unlike ccsr’s separate .md finding-exclusion (which is written to the results JSON), this pre-model strip records nothing on either deliverable, the pull-request comment or the results file. Whether the model, which still sees the bare filename, mentions the file in its report depends on the model, and we did not run that step.

The one panel member whose drop is \textbf{disclosed} is VVAH, in its own report.

We ran each harness’s own selection code on the one file, without the model, with controls to separate potentially overlapping reasons: \nohy{src/handler} and \nohy{tests/handler.py} in earlier runs, and the plain \nohy{src/handler.py} version in the final run described next.

All six were run on the fixture shown above, with the \nohy{runpy.run\_path} loader. Selection does not depend on the loader. The loader matters for a different reason. An earlier version loaded the file with exec(open(p).read()).

A Semgrep check, using a local rule modeled on the standard exec rule, flagged that line at error level on \nohy{app.py,} a file every harness selects, pointing straight at the hidden file. With \nohy{runpy.run\_path,} the same check still reports a warning on \nohy{app.py,} but only a lower-severity dynamic-import one. ai-deep-sast was re-run under tree-sitter 0.26 and tree-sitter-python 0.25, newer than those in its original run. Its drop happens at file collection, which does not use the grammar.

\textbf{\nohy{RepoAudit}} and \textbf{ai-deep-sast} drop purely on the missing extension (the \nohy{tests/handler.py} control is scanned). \nohy{OpenAnt’s} extension check runs before its test-folder check, so the extensionless file is dropped there before any skip count can be written.

\textbf{Codex Security} drops on the extension. Its ranking step has a separate exclusion of test directories \nohy{(generate\_\allowbreak{}rank\_\allowbreak{}input.\allowbreak{}py:\allowbreak{}43-85} and :283 at 455d7c8). VVAH’s drop is the \nohy{tests/} directory and not the extension \nohy{(src/handler} is kept).

\section{Why harnesses miss vulnerabilities}

Each drop in Section 4 is traced to the harness’s source or shipped instructions at a specific commit. The grouping below is our suggested classification of those mechanisms, and some drops fit more than one group.

For \nohy{OpenAnt,} the reasons for its design choices are stated by its authors (Korda \& Evron, 2026), and we know them for raptor, which we maintain. For the other harnesses, we infer them from code and documentation.

Each subsection names the design principle that the drops violate.

\subsection{The harness reads code differently from how it runs}

Principle: a harness reads code the way the runtime does, and treats code it cannot classify as unresolved, never as absent.

This is the gap Ptacek and Newsham described for intrusion detection, where the monitor misses input that the target processes. They studied it as an attack, while most drops here occur without one. Sections 5 to 7 show the attack case. Almost all of these drops are in the harness’s own code, often as string checks on a file suffix, a path or a name, which never ask how the runtime will treat the code.

\begin{itemize}

\item Identity from names and text. A file’s language is decided by its extension, so a file with no extension is never read \nohy{(OpenAnt,} Codex, \nohy{RepoAudit,} ai-deep-sast, \nohy{ToB,} deepsec). A path containing “log” is skipped, which includes \nohy{auth/login.py} (Rivian’s ai-sast). A user function named like a PHP built-in is resolved as the built-in, which removes its only incoming call \nohy{(OpenAnt).}

\item Unrecognized code treated as absent. \nohy{OpenAnt} identifies entry points by decorators, parameter patterns and framework conventions (Korda \& Evron, 2026). Handlers outside that list, such as Celery tasks or AWS Lambda handlers, are pruned as unreachable. The \nohy{OpenAnt} paper’s limitations section does not list this gap, or the others we report for \nohy{OpenAnt.} We report them here as gaps in our own published design. Macro bodies that the parser cannot read and code that the build does not compile are dropped the same way.

\end{itemize}

These rules were written to describe typical code. Applied as filters, they decide what counts as code. This is what we mean by descriptive harnesses becoming prescriptive.

\subsection{The harness takes instructions from the code it checks}

Principle: text in the scanned code is data and never an instruction.

\begin{itemize}

\item \textbf{Markers read by deterministic steps.} \# nosemgrep removes a finding (ai-deep-sast). @generated anywhere in a file’s diff removes the whole file (ccsr). A \nohy{tests/} folder removes the file from analysis, with notice, in \nohy{OpenAnt} and VVAH.

\item \textbf{Text read by the model.} A comment claiming that code is safe can lead the model to dismiss a real vulnerability.\newline \nohy{RepoAudit} dropped a real, reachable finding because of a planted safety comment in one of three runs. Comments that argue code is safe had no significant overall effect on detection in one study of eight models, although the same study reports 28 individual induced misses (Thornton, 2026, Section 5.2, Table 4, and Section 5.5).

In another, a comment containing a fake verdict in the detector’s expected output format turned 55–99\% of correctly detected vulnerabilities into “benign” results, without any optimization, a baseline the authors place outside their threat model (Sun, Oprea \& Wong, 2026, Section 5.5 and Figure 2).

The two studies used different languages, models and benchmarks, and each supports only its own half of the contrast. Our reading of the two is that text aimed at how the harness parses the model’s output did more than text arguing that the code was safe.

\item \textbf{Prompt injection. }Instructions planted in code or pull-request text can redirect the harness’s agent. This has been shown against AI code review agents, including Claude Code Security Review (Guan, Liu \& Zhong, 2026). The commit we studied, 0c6a49f of 11 February 2026, predates that public disclosure.

\end{itemize}

Codex Security follows this principle: it treats repository text as data and suppresses a finding only on evidence from the source. raptor wraps untrusted input and reopens findings the model was talked out of.

\subsection{The harness loses results it already had}

Principles: a failure is not reported as a pass. “Not analyzed” is a separate result from “clean”. Stored results are keyed on everything that can change them.

Saltzer and Schroeder (1975) state the fail-safe default for access decisions: base them on permission rather than exclusion. Their reason is that a mistake in a mechanism that excludes access tends to fail by allowing access, and that failure may go unnoticed in normal use. For a harness, the protected property is coverage, so the safe default is to analyze or flag, and every skip needs a stated reason.

\begin{itemize}

\item \textbf{Errors and limits recorded as “no vulnerability.”} A model reply that does not start with “TRUE” is read as safe (Rivian’s ai-sast). An unparseable reply becomes an empty result (iris). An answer the regular expression cannot match becomes no finding \nohy{(RepoAudit).} Caps on the number of units \nohy{(OpenAnt} -{}-limit) and on fetched files (ccsr, in a fallback mode) drop the remainder. Code truncated to fit the model’s context window can be lost the same way (Korda \& Evron, 2026). Ptacek and Newsham describe a related case, in which the monitor’s resources are exhausted.

\item \textbf{Stored results matched on the wrong key.} A new finding whose type and title match an older record marked “fixed” is merged into it, and the old status wins (deepsec). A cached verdict is keyed without the harness version and never expires (raptor). A repository is skipped because its output folder already contains a results file \nohy{(VulnHunter).}

\item \textbf{Rules that never run.} raptor ships documented \nohy{CodeQL} queries that are never added to the set a scan runs, and VVAH has a complete deserialization analyzer that no shipped profile turns on (class 9). The report gives no indication that these checks were not run.

\item \textbf{Skips recorded where users do not look.} Examples are console or stderr output (Rivian’s ai-sast, \nohy{OpenAnt’s} extension drop), a run log (raptor’s verification gate), or a total count with no item named \nohy{(OpenAnt’s} PHP path).

\item \textbf{Engine defaults left on.} A Semgrep wrapper that does not pass -{}-disable-nosem inherits Semgrep’s handling of \# nosemgrep (ai-deep-sast). A Semgrep skill with no ignore override inherits Semgrep’s default skips of minified and vendored files \nohy{(ToB).} \nohy{CodeQL’s} build-based extraction misses code the build does not compile \nohy{(ToB,} iris).

\end{itemize}

\subsection{How the steps combine}

Principles: any single check can raise a finding. Protection is the same across languages and modes.

\begin{itemize}

\item \textbf{One step is enough.} A finding must pass every step, so a single dropping step removes it. In Section 6, stacking three placements on one file also means the two notices that the test-folder and diff drops each write on their own are never written.

\item \textbf{Requiring agreement drops findings.} raptor demotes a finding that no engine other than the model corroborates. Its built-in prefilter covers C, C++ and Python, so a correct finding in PHP is dropped unless another engine or a domain-model invariant backs it, while a less certain “suspicious” finding would have been reported.

\item \textbf{Changes treated as local.} In pull-request mode, deepsec selects only added, modified, renamed or copied files. A deleted file that held a check, or an unchanged file made unsafe by a change elsewhere, is never analyzed.

\item \textbf{Protection varies by language and mode.} \nohy{OpenAnt’s} Python path resolves a same-file function over the built-in and names pruned units in \nohy{pruned\_units.json,} while its PHP path does neither. Drops also concentrate in secondary modes: pull-request scanning, resume, fallback and opt-in channels.

\item \textbf{Noise filters create unreported areas.} Steps added to reduce false positives define areas where findings are not counted: test folders, generated files, documentation files, findings matched to “false positive” records, and uncorroborated findings. Code placed in those areas is not analyzed, and in several harnesses the skip is not reported.

\end{itemize}

\subsection{Summary}

\begin{center}\small
\begin{tabularx}{1.000\textwidth}{>{\raggedright\arraybackslash\hsize=0.814\hsize}X>{\raggedright\arraybackslash\hsize=0.270\hsize}X>{\raggedright\arraybackslash\hsize=1.917\hsize}X}
\toprule
\textbf{Principle} & \textbf{Section} & \textbf{Drop classes} \\
\midrule
Read code the way the runtime does, and treat unclassified code as unresolved & 8.1 & 1 Unrecognized File Types, 2 Argument Count Mismatch, 3 Built-in Name Conflict, 4 Unrecognized Entry Points, 11 Keyword-Based Skipping (path keywords), 12 Unparsed Code Blocks \\
Text in the scanned code is data & 8.2 & 8 Suppression Comments, 11 Keyword-Based Skipping (@generated, \nohy{tests/),} content-steered verdict and prompt injection (Section 3, unnumbered) \\
A failure is not reported as a pass & 8.3 & 6 Scan Limit Cutoff, 7 Verification Filtering (unparseable replies) \\
“Not analyzed” is separate from “clean” & 8.3 & 6 Scan Limit Cutoff, 7 Verification Filtering, 9 Inactive Rules \\
Key stored results on everything that can change them & 8.3 & 5 Overridden Findings, 10 Stale Caching \\
Any single check can raise a finding & 8.4 & 7 Verification Filtering (raptor’s corroboration gate) \\
Protection is the same across languages and modes & 8.4 & 3 Built-in Name Conflict (PHP vs. Python), 6 Scan Limit Cutoff (ccsr fallback), 8 Suppression Comments (raptor’s opt-in channel), 10 Stale Caching \nohy{(VulnHunter} resume), 1 Unrecognized File Types and 13 Diff Blindspots in pull-request modes \\
\bottomrule
\end{tabularx}
\end{center}

\section{Building, testing and using harnesses}

\subsection{Building harnesses}

These recommendations apply with or without an attacker. Each addresses findings from Section 8.

\begin{itemize}

\item Report every skipped file, function or finding in the report itself, with the reason. Several drops in Section 8.3 were recorded only in console output, a log or a total count.

\item Keep uncertain code in scope at a lower priority, and mark unresolved calls and entry points as unresolved. This addresses the reachability drops in Section 8.1. Under a fixed budget it still loses coverage, but the loss becomes visible.

\item Record errors, timeouts and unreadable model replies as “not analyzed.” This addresses the parsing drops in Rivian’s ai-sast, iris and \nohy{RepoAudit.}

\item Key cached and stored results on the harness version and on everything else that can change the result, and reopen findings that reappear after being marked fixed. This addresses deepsec, raptor and \nohy{VulnHunter.}

\item Let a finding stand without a second engine’s agreement, and use agreement to rank findings. This addresses raptor’s verification gate.

\item Override engine defaults that suppress or skip code, and record any suppression that remains. This addresses ai-deep-sast and \nohy{ToB.}

\item Test that every shipped rule and analyzer runs in at least one default configuration. This addresses the inactive rules in raptor and VVAH (class 9).

\item In pull-request mode, include deleted files and re-analyze callers of changed symbols. This addresses deepsec (class 13).

\item Separate target text from instructions in every model prompt, and never let a comment alone close a finding. This addresses the drops in Section 8.2.

\end{itemize}

\nohy{OpenAnt’s} model stages already apply several of these principles: they have an “inconclusive” verdict, send all detection results, including “safe” and “protected,” on to verification, and require several exploitation attempts before ruling a finding out (Korda \& Evron, 2026). Its static stages, which remove 96.4\% of functions before any model runs, report some drops but have no mechanism that keeps uncertain code in scope.

The same protections can be added there, for example an “unresolved” state for calls and entry points, or a model check on a sample of pruned code. Because these drops happen before or after the model runs, upgrading the model does not remove them.

\subsection{Harnesses that check code written by others}

When the code’s author may be hostile, every rule in Sections 8.1 and 8.2 is a rule the author can satisfy or avoid. Open-source harnesses publish these rules, and the \nohy{OpenAnt} paper lists its entry-point patterns. The steps that select code need the same treatment as text given to the model, because their input is untrusted.

Ptacek and Newsham noted that a monitor can resolve such ambiguities only with information about how the end system behaves. Later work removed the ambiguities from the traffic before the monitor saw it (Handley, Paxson \& Kreibich, 2001).

For a harness, the first approach means resolving files, calls and entry points the way the runtime does. Running several harnesses helps less when they share the same rules: in Section 7.2, four of six harnesses dropped the same file at selection because it had no extension.

\subsection{Testing for missed findings}

The demonstrations in Sections 5 to 7 are tests of one kind: change the code in a way that does not change its behavior, and check that the finding stays the same. Section 7.2 compares the hidden file with controls \nohy{(src/handler.py} in the final run, and \nohy{src/handler} and \nohy{tests/handler.py} in earlier runs) to separate the reasons for each drop. Section 6 reasons through undoing one placement at a time.

Builders can run the same tests on their own harnesses by varying:

\begin{itemize}

\item File names and extensions

\item Folders

\item Function names

\item Comments and markers

\item The scan mode

\end{itemize}

These tests need no labeled dataset, only the original and the changed version. The \nohy{OpenAnt} paper avoids public benchmarks partly because their content may appear in model training data (Korda \& Evron, 2026).

Changed versions are generated fresh, and the deterministic steps involve no model at all, so that concern does not apply to them. The same method is established for testing static analyzers (Zhang et al., 2023).

\subsection{Using harnesses}

\begin{itemize}

\item Read a clean report as “no finding was produced.” In Section 7.2, five of six harnesses dropped the file before the model saw its code, without notice.

\item Look for skips beyond the main report. Some harnesses record them only in console output or logs (Section 8.3).

\item If you scan only pull requests, also run full scans. Pull-request modes drop deleted and unchanged files (Section 8.4).

\item Clear caches and rerun after updating a harness. Some harnesses reuse verdicts from earlier versions (Section 8.3).

\item When scanning code written by others, assume that file names, folders and comments may have been chosen with the harness in mind (Sections 5 to 7).

\end{itemize}

\section{Limitations}

\begin{itemize}

\item This work started as hardening of the two harnesses we maintain. We examined \nohy{OpenAnt} and raptor most closely, and decided to use them most in examples. We examined the other harnesses to establish how widely the problem appears.

\item The 12 harnesses were selected for illustration. This is not a ranking.

\item We did not measure how often these cases occur in real repositories.

\item Counting silent entries only, six classes rest on a single harness (2, 3, 5, 6, 8, 13) and three more on two (4, 9, 10). A class describes a mechanism and makes no assertions about prevalence.

\item The Section 7 demonstrations exercised each harness’s file-selection stage without its model. Whether a model with file tools would recover the dropped file was not tested.

\item Design reasons are stated for \nohy{OpenAnt} (Korda \& Evron, 2026), known to us for raptor, and inferred for the other harnesses.

\item We tested security harnesses only. Similar failures are reported elsewhere: answer extraction in LLM evaluation frameworks (Yu et al., 2025) and document retrieval in retrieval-augmented generation (Barnett et al., 2024). This supports our expectation that the reliability findings apply beyond security, but we did not test it.

\end{itemize}

\section{Disclosure}

None of the drop points we identified is a vulnerability as defined for the purposes of coordinated disclosure.

Each is an engineering choice, or an ordinary bug, in how a harness selects, parses, verifies, or reports, and none is a vulnerability in the harness’s own code, such as code execution or data theft through the harness. (The prompt-injection case cited in Section 8.2 is a third-party disclosure, already public.)

We therefore did not treat our findings as security disclosures, but as a matter of professional courtesy we nevertheless informed the maintainers of every harness named in this paper before publication, with the file, line and commit for each item.

Several maintainers responded, and many of the issues discussed here were fixed before this paper was finalized.

The two harnesses we maintain carry the most entries. Those items have since been fixed in \nohy{OpenAnt} and raptor. Rivian also responded, confirming their own fixes.

\section{Conclusion}

We started by hardening the two harnesses we maintain and found that the rules governing how harnesses work around the model can drop real, reachable vulnerabilities.

We then examined how far the problem goes, mapping 13 points where this happens across 12 open-source harnesses. The drops do not depend on the model, and the report usually gives no sign of them.

With no attacker, they leave vulnerabilities undetected. An attacker who knows a harness’s rules can use the same drops to hide a vulnerability. In our demonstrations, one file was dropped before the model saw its code by six harnesses at once, five of them without notice. The model can also be steered by comments or by instructions planted in the code.

Most of the drops happen in deterministic steps added to control cost and false positives. Harnesses built to describe typical code end up prescribing what is analyzed, and code that does not conform is treated as absent.

Many of these drops follow the gap Ptacek and Newsham described for intrusion detection, where the monitor reads input differently from the host it protects.

Other drops come from a harness reporting its own failures as clean results, which a fail-safe default, adapted from Saltzer and Schroeder, rules out.

Because harness rules are often public, harness design needs adversarial review: builders should ask how an attacker could use each design choice against the people who rely on the harness.

Section 8 gives principles for building resilient harnesses, meaning harnesses that are accurate about what they analyzed and harder to evade.

Section 9 turns them into recommendations. Builders should keep uncertain code in scope, record failures as “not analyzed,” report what was skipped, treat text in the code as data, and test with changes that do not alter the code’s behavior. Users should read a clean report as “no finding was produced.”

The reliability findings likely apply to other LLM harnesses. The evasion findings apply wherever an attacker can write or influence the code being checked.

\section*{Acknowledgments}

\addcontentsline{toc}{section}{Acknowledgments}

We thank Thomas H. Ptacek, John Cartwright, Daniel Cuthbert, Michal Kamensky, Daniel Goldberg, Yoni Rosenshein, and Kat Traxler.

\section*{References}
\addcontentsline{toc}{section}{References}
\begingroup\small\raggedright\setlength{\parskip}{0.45em}\everypar{\hangindent=1.5em\hangafter=1}

Barnett, S., Kurniawan, S., Thudumu, S., Brannelly, Z., \& Abdelrazek, M. (2024). Seven failure points when engineering a retrieval augmented generation system. CAIN 2024. \url{https://arxiv.org/abs/2401.05856}

Free Software Foundation. \nohy{(n.d.).} Diagnostic pragmas. In Using the GNU Compiler Collection (GCC). Retrieved October 2026, from \url{https://gcc.gnu.org/onlinedocs/gcc/Diagnostic-Pragmas.html}

\nohy{GitHub.} \nohy{(n.d.-a).} Preparing your code for \nohy{CodeQL} analysis. \nohy{GitHub} Docs. Retrieved October 2026, from \url{https://docs.github.com/en/code-security/codeql-cli/getting-started-with-the-codeql-cli/preparing-your-code-for-codeql-analysis}

\nohy{GitHub.} \nohy{(n.d.-b).} Using pagination in the REST API. \nohy{GitHub} Docs. Retrieved October 2026, from \url{https://docs.github.com/en/rest/using-the-rest-api/using-pagination-in-the-rest-api}

Guan, A., Liu, Z., \& Zhong, G. (2026). Comment and Control: Prompt injection to credential theft in Claude Code, Gemini CLI, and \nohy{GitHub} Copilot Agent. Coordinated disclosure, 15 April 2026. \url{https://oddguan.com/blog/comment-and-control-prompt-injection-credential-theft-claude-code-gemini-cli-github-copilot/}

Guo, J., Wang, C., Xu, X., Su, Z., \& Zhang, X. (2025). \nohy{RepoAudit:} An autonomous LLM-agent for repository-level code auditing. ICML 2025. \url{https://arxiv.org/abs/2501.18160}

Handley, M., Paxson, V., \& Kreibich, C. (2001). Network intrusion detection: Evasion, traffic normalization, and end-to-end protocol semantics. 10th USENIX Security Symposium, 115–131. \url{https://www.usenix.org/conference/10th-usenix-security-symposium/network-intrusion-detection-evasion-traffic-normalization}

Korda, N., \& Evron, G. (2026). \nohy{OpenAnt:} LLM-powered vulnerability discovery through code decomposition, adversarial verification, and dynamic testing. \nohy{arXiv:2606.19149.} \url{https://arxiv.org/abs/2606.19149}

Li, Z., Dutta, S., \& Naik, M. (2025). IRIS: LLM-assisted static analysis for detecting security vulnerabilities. ICLR 2025. \url{https://arxiv.org/abs/2405.17238}

Linux man-pages project. \nohy{(n.d.).} execve(2). Linux manual pages. Retrieved October 2026, from \url{https://man7.org/linux/man-pages/man2/execve.2.html}

LLVM Project. \nohy{(n.d.).} FAQ and how to deal with common false positives. Clang Static Analyzer documentation. Retrieved October 2026, from \url{https://clang.llvm.org/docs/analyzer/user-docs/FAQ.html}

Ptacek, T. H., \& Newsham, T. N. (1998). Insertion, evasion, and denial of service: Eluding network intrusion detection. Secure Networks, Inc. \url{https://www.icir.org/vern/Ptacek-Newsham-Evasion-98.pdf}

Saltzer, J. H., \& Schroeder, M. D. (1975). The protection of information in computer systems. Proceedings of the IEEE, 63(9), 1278–1308. \url{https://doi.org/10.1109/PROC.1975.9939}

Semgrep. \nohy{(n.d.-a).} CLI reference. Semgrep Docs. Retrieved October 2026, from \url{https://docs.semgrep.dev/cli-reference}

Semgrep. \nohy{(n.d.-b).} Ignore files, folders, and code. Semgrep Docs. Retrieved October 2026, from \url{https://docs.semgrep.dev/ignoring-files-folders-code}

Sun, L., Oprea, A., \& Wong, E. (2026). Syntax- and compilation-preserving evasion of LLM vulnerability detectors. \nohy{arXiv:2602.00305.} \url{https://arxiv.org/abs/2602.00305}

Taylor, R. N., Medvidovic, N., \& Dashofy, E. M. (2010). Software architecture: Foundations, theory, and practice. Wiley.

Thornton, S. (2026). Can adversarial code comments fool AI security reviewers: Large-scale empirical study of comment-based attacks and defenses against LLM code analysis. \nohy{arXiv:2602.16741.} \url{https://arxiv.org/abs/2602.16741}

Yu, Q., Zheng, Z., Song, S., Li, Z., Xiong, F., Tang, B., \& Chen, D. (2025). \nohy{xFinder:} Large language models as automated evaluators for reliable evaluation. ICLR 2025. \url{https://arxiv.org/abs/2405.11874}

Zhang, H., Pei, Y., Chen, J., \& Tan, S. H. (2023). Statfier: Automated testing of static analyzers via semantic-preserving program transformations. \nohy{ESEC/FSE} 2023. \url{https://doi.org/10.1145/3611643.3616272}

\endgroup

\end{document}